%% file: pairseed-arxiv.tex
\documentclass[10pt]{article}
\usepackage[nofonts]{dgleich-common}

\usepackage[T1]{fontenc}
\RequirePackage[scaled=0.8]{helvet}%
\RequirePackage[scaled=0.75]{beramono}%
\def\lstbasicfont{\fontfamily{fvm}\selectfont}
\makeatletter  
\setboolean{@dgleich@sanssmallcaps}{true}
\makeatother

\usepackage{dgleich-math}
\usepackage{dgleich-tensor}

\usepackage{graphicx} 
\usepackage{caption}
\usepackage{mwe}

\definecolor{mygreen}{rgb}{0.6, 0.7607843137254902, 0.30196078431372547}

\usepackage{enumitem}
\usepackage{calc}
\usepackage{amssymb}
\newcommand*\arc{{\fontfamily{pbk}\fontseries{db}\selectfont+}}
\usepackage{tabularx}
\usepackage{booktabs}
\usepackage{algorithmic}
\usepackage[ruled,vlined]{algorithm2e}
\newcommand{\hide}[1]{}
\newcommand{\xhdr}[1]{\vspace{0.25mm}{\bf #1.}\hspace{0.5mm}}
\graphicspath{{figures/}}

\usepackage{subcaption}

\newcommand{\phantomsubfigure}[1]{\begin{subfigure}[b]{0.1\textwidth}\phantomcaption\label{#1}\end{subfigure}}

\usenatbib
\usehyperref

\usepackage[most]{tcolorbox}
\definecolor{plots1}{rgb}{0.0,0.605603,0.97868}
\definecolor{plots2}{rgb}{0.888874,0.435649,0.278123}
\definecolor{plots3}{rgb}{0.242224,0.643275,0.304449}
\definecolor{plots4}{rgb}{0.76444,0.444112,0.824298}
\definecolor{plots5}{rgb}{0.675544,0.555662,0.0942343}
\definecolor{plots6}{rgb}{0.00000482118,0.665759,0.680997}
\definecolor{plots7}{rgb}{0.930767,0.367477,0.57577}

\graphicspath{{./}{./figures/}}

\usepackage{enumitem}

\usenatbib
\usehyperref

\makeatletter
\def\blfootnote{\xdef\@thefnmark{}\@footnotetext}
\makeatother

\title{Pairwise Link Prediction}
\author{Huda Nassar, Austin R.~Benson, David F.~Gleich}
\date{}

\begin{document}
\marginnote[500pt]{\fontsize{7}{9}\selectfont
Huda Nassar, Purdue University\\
\url{hnassar@purdue.edu}\\
\noindent Austin R.~Benson, Cornell University\\
\url{arb@cs.cornell.edu}\\
\noindent David F.~Gleich, Purdue University\\
\url{dgleich@purdue.edu}
}


\maketitle
\input{000-abstract}

\input{001-introduction}
\input{002-related}
\input{003-methods}
\input{004-evaluations}
\input{005-results}

\input{006-linkpred}
\input{007-conclusion}

\section*{Acknowledgements} 
Supported by NSF IIS-1422918, IIS-1546488, CCF-1909528, NSF Center for Science of Information STC, CCF-0939370, NASA, Sloan Foundation,
DARPA SIMPLEX, NSF DMS-1830274,
ARO W911NF-19-1-0057, and
ARO MURI.

\bibcolumns=3
\begin{fullwidth} 
\bibliographystyle{dgleich-bib}
\bibliography{sample-bibliography}
\end{fullwidth}

\end{document}

%% file: 000-abstract.tex

\begin{abstract}
  Link prediction is a common problem in network science that transects many
  disciplines. The goal is to forecast the appearance of new links or to
  find links missing in the network.
  Typical methods for link prediction use the topology of the network to predict
  the most likely future or missing connections between a pair of nodes.
  However, network evolution is often mediated by higher-order structures
  involving more than pairs of nodes; for example, cliques on three nodes (also
  called triangles) are key to the structure of social networks, but the
  standard link prediction framework does not directly predict these structures.
  To address this gap, we propose a new link prediction task called
  ``pairwise link prediction'' that directly targets the prediction of new triangles, where
  one is tasked with finding which nodes are most likely to form a triangle with a given
  edge.
  We develop two PageRank-based methods for our pairwise link prediction problem
  and make natural extensions to existing link prediction methods.
  Our experiments on a variety of networks show that diffusion based methods are 
  less sensitive to the type of graphs used and more consistent in their results.
  We also show how our pairwise link prediction framework can be used to get
  better predictions within the context of standard link prediction evaluation.
\end{abstract}


%% file: 001-introduction.tex
\section{Introduction}
Networks are a standard tool for data analysis in which links between data
points are the primary object of study.
A fundamental problem in network analysis is \emph{link prediction}~\cite{LibenNowell-link-prediction-2007,Lu-linkpred-survey-2011},
which is typically formulated as a problem
of identifying pairs of nodes that will either form a link in the future (when
viewing the network as evolving over time) or whose
connection is missing from the data~\cite{Clauset-hierarchial-2008}.
The link prediction problem has applications in a variety of domains.
For instance, in online social networks of friendships, predicting that two
people will form a connection can be used for friendship
recommendation~\cite{Backstorm-linkprediction-2010}.
Similarly, predicting new links between users and items on platforms such as
Amazon and Netflix can be used for product
recommendation~\cite{Gomez-Uribe-netflix-2016}.
And in biology, link prediction is used to identify novel interactions between
genes, diseases, and drugs within interaction
networks~\cite{Hsu-multimodal-2018}.
In the settings above, the link prediction problem is oriented around---and
evaluated in terms of---the identification of \emph{pairs of nodes} that are
likely to be connected. However, there is mounting evidence that the organization
and evolution of networks is centered around higher-order interactions involving
more than two nodes~\cite{Milo-motif-2002,Milo-superfamilies-2004,Benson-organization-2016,Benson-simplical-2018,Lambiotte-models-2019}.
In the case of social networks, triangles (cliques on three nodes) are extremely common
due to various sociological mechanisms driving triadic
closure~\cite{Easley-book-2010,Holland-structure-1977,Granovetter-ties-1977,Rapoport-transitivity-1953}.
Methods for link prediction are indeed motivated by these ideas.
For instance, the Jaccard similarity between the sets of neighbors of two nodes---a common heuristic
for link prediction~\cite{LibenNowell-link-prediction-2007}---measures the number of triangles
that would be created if the two nodes are linked, normalized by the total number of neighbors
of the two nodes. 
Still, such methods are used to make predictions on
\emph{pairs of nodes}, rather than a prediction on the appearance of the
higher-order structures directly.
Here, we develop a framework for directly predicting the appearance of
a higher-order structure.  We focus on the case of triangles, which is one of the
simplest higher-order structures while also being critical to social network
analysis.  Again, classical link prediction is centered around the following
question: given a node $u$ in the network, which nodes are likely to link to
$u$? This scenario is illustrated in Figure~\ref{fig:pairwise_A}.
Our framing of the problem is similar, but we instead ask the
following: given an edge $(u,v)$ in the network, which nodes are likely to
connect to both $u$ and $v$?  We call this the \emph{pairwise link prediction problem},
and it is illustrated in Figure~\ref{fig:pairwise_B}.
There are several scenarios where the pairwise link prediction problem is
natural, such as recommending a new friend to a couple on an online social
network, recommending a movie to a couple in a video site, or predicting an
effective drug given a disease-gene pair.  
We devise two new algorithms for the pairwise link prediction problem.
The first is based on a variant of seeded (personalized) PageRank that uses
multiple seeds, namely, one seed at each end point of the edge for which we
are trying to predict new triadic connections. The second is based on a PageRank-like
iteration that puts more weight on edges that participate in many triangles. In this
sense, the method reinforces triangles, and we call the method ``Triangle
Reinforced PageRank'' (TRPR). We compare these algorithms to natural extensions
of local similarity measures that are common in link prediction, such
as Jaccard similarity~\cite{LibenNowell-link-prediction-2007},
Adamic-Adar similarity~\cite{Adamic-friends-web-2003}, and
preferential attachment~\cite{Newman-growing-networks-2001}. 
For a given edge, each of the above methods produces a score for the remaining
nodes in the graph. We find that our
proposed diffusion based methods are the least sensetive to the graph type and degree distribution and often produce the top results.
We provide code for all the methods used in this paper in the repository:
\[
\text{\url{https://github.com/nassarhuda/pairseed}}
\]
\hide{
outperform the baseline measures
based on local neighborhood information on a number of synthetic benchmark and
real-world datasets. For instance, on predicting future links in a temporal graph,
our methods had median
AUC scores of 0.93 compared with 0.76 for the baseline methods. 
Based on this success of our PageRank-based methods for pairwise link
prediction, we then go back and adapt them for the classical link prediction
problem.
We find that these adapted methods out-perform traditional seeded PageRank
on a number of real-world datasets with average AUC increases of up to 0.28 in
predicting missing drug interactions.}
\begin{marginfigure}
 \phantomsubfigure{fig:pairwise_A}
 \phantomsubfigure{fig:pairwise_B}  
 \includegraphics[width=\linewidth]{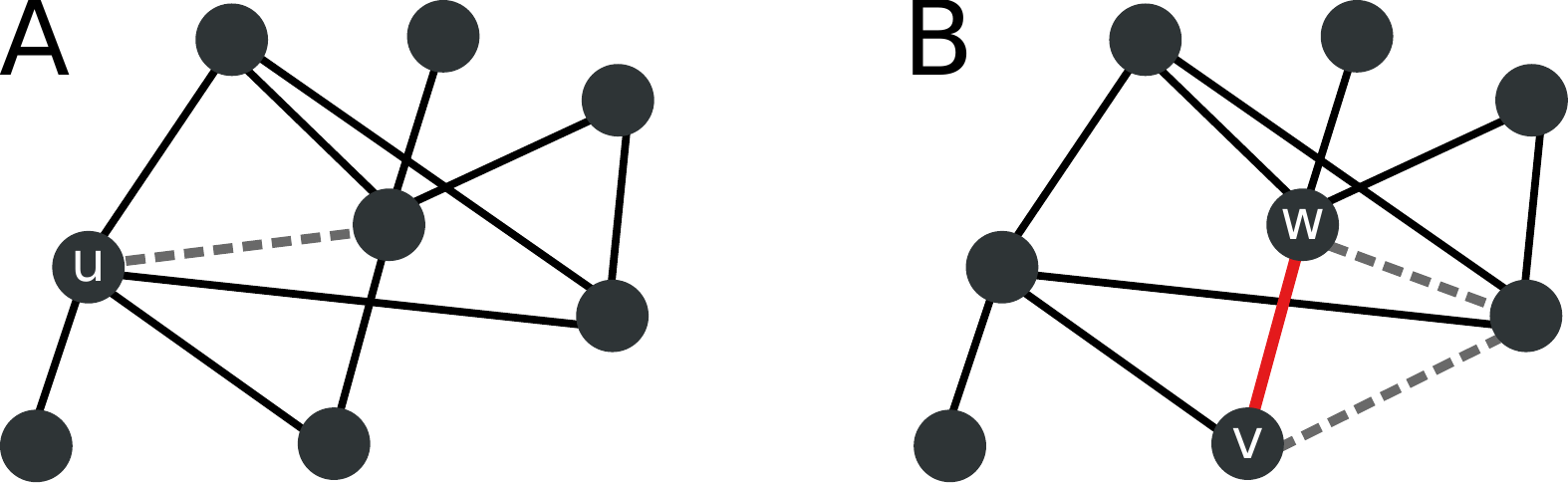}
 \caption{\textbf{(A)} In standard link prediction, we are tasked with finding nodes that are likely to link to a given node $u$.
   \textbf{(B)} In this paper, we study pairwise link prediction, where we are tasked with finding nodes that are likely
   to form a triangle with a given edge $(v, w)$.}
 \label{fig:pairwise}
\end{marginfigure}

%% file: 002-related.tex
\section{Background and Related Work}
\label{sec:related}
We now briefly review some related work in link prediction and higher-order
structure.  As part of this, we will go over methods that we will generalize in
the next section for the pairwise link prediction problem.  All of these methods
assign some similarity score between pairs of nodes, where a larger similarity
is indicative of pairs that are likely to connect.  For notation, we use
$\Gamma(u)$ to denote the set of neighbors of node $u$ in the graph.

\subsection{Local methods}
Several approaches to link prediction are based on local information in the graph,
namely a score is assigned to a pair of nodes $w$ and $u$ based on their 1-hop
neighborhoods $\Gamma(w)$ and $\Gamma(u)$.
One approach that falls under this category stems from the idea that as
$\lvert \Gamma(w)\cap\Gamma(u) \rvert$ increases, the chance that $u$ and $v$ are connected
also increases~\cite{Newman-growing-networks-2001}.
Here, $\lvert \Gamma(w)\cap\Gamma(u) \rvert$ is the number of triangles that would be formed
if $u$ and $v$ were connected. Often, this number is normalized by the size of the neighborhoods,
which gives rise to the Jaccard similarity between two nodes $w$ and $u$:
\[
\frac{\lvert \Gamma(w)\cap\Gamma(u) \rvert}{\lvert \Gamma(w)\cup\Gamma(u) \rvert}.
\]
The Adamic--Adar similarity measure~\cite{Adamic-friends-web-2003} is a local
score that assigns similarity between two nodes based on how important their
common neighbors are, where importance is measured by the degree of a given
node. Formally, the Adamic--Adar similarity measure between nodes $u$ and $v$
is:
\[
\sum_{z \in \Gamma(w) \cap \Gamma(u)} \frac{1}{\log(\lvert \Gamma(z) \rvert)}.
\]
A third local method is based on preferential attachment, where nodes are more likely to
connect to \textit{established} nodes in the network, and, \textit{established}
nodes have a higher chance to connect to each
other~\cite{Barbasi-preferential-attachment-1999,Newman-growing-networks-2001}.
Using degree as a proxy for how established a node is, the preferential attachment
score between nodes $w$ and $u$ is:
\[
\lvert \Gamma(w) \rvert \cdot \lvert \Gamma(u) \rvert.
\]

\subsection{Global methods}
Another set of approaches for link prediction are based on aggregating (weighted
or normalized) path counts of varying lengths. In contrast to the local methods
described above, these methods use global information about the entire
network. For example, the Katz similarity counts the number of paths between two
nodes, weighting paths of length-$k$ by
$\beta^k$~\cite{Katz-1953,LibenNowell-link-prediction-2007}.  Another class of
global methods are methods based on conservative diffusions such as
PageRank~\cite{Page-PageRank-1999}.  Such diffusion methods are typically
\emph{seeded} by a particular node $u$, and the similarity of $u$ to all other
nodes is given by the amount of ``mass'' that diffuses to each other node.  We
will make use of PageRank-like methods in the next section.

\subsection{Higher-order structure}
Since a network encodes pairwise relationships (edges) between elements (nodes),
the link prediction problem is natural in many cases. Nevertheless, recent
studies have shown that networks evolve through higher-order interactions, i.e.,
much of the structure in evolving networks involves interactions between more
than just two nodes~\cite{Benson-simplical-2018}. Recent research has also introduced the problem of predicting the
\textit{time} when an edge addition will close a triangle~\cite{Dave-TCTP-2019}.
Furthermore, random graph models
constructed from distributions of triangles have shown to be good fits for
real-world data~\cite{Eikmeier-hyperkron-2018}, providing additional evidence
that triadic relationships are important to the assembly of networks.


%% file: 003-methods.tex
\section{Methods}
We propose several methods for the pairwise link prediction problem.
First, we extend the three local methods described above to measure
node-edge similarity. 
After, we propose diffusion-based methods akin to seeded PageRank.

\subsection{Local similarity measures for pairwise prediction}
\label{sec:local}
Our goal here is to extend common local methods for link prediction to the
scenario of pairwise link prediction. In other words, instead of computing
similarity between nodes, we now compute similarity between an edge and a node.
To do this, we simply replace the neighborhood of one node with the neighborhood
of an edge. This requires that we specify what the neighborhood of an edge $(u,v)$
should capture.  We define:
\begin{align*}
  \Gamma((u,v)) &= \{\text{node } z \;\vert\; z \text{ is connected to either or both nodes } u, \text{and }v\} \\
                &= \Gamma(u) \cup \Gamma(v) \setminus \{u,v\}.
\end{align*}
Note that this is akin to the boundary of a set of vertices in the graph that
is often used to define the size of a cut, which --- for an edge --- would correspond to the
union of neighborhoods. This definition is contrary to what we use in~\cite{Nassar-pairwise-2019}, where we define the neighborhood to be the intersection. Initially, the intersection of neighborhoods was a natural choice, but in practice the intersection set is very limiting specially in scenarios when an edge is connected to the rest of the graph, yet does not participate in any triangles. We should still be able to make predictions on such edges, and our choice of the union of neighborhoods handles this.

Using the substitution gives us three similarity measures that will
compute the similarity of an edge to a node.
\begin{itemize}
\item Jaccard Similarity (JS).
\begin{flalign*}
\text{JS}(w,(u,v)) &= \frac{\lvert \Gamma(w) \cap \Gamma((u,v)) \rvert}{\lvert \Gamma(w) \cup \Gamma((u,v)) \rvert} &
\end{flalign*}
\item Adamic--Adar (AA).
\begin{flalign*}
\text{AA}(w,(u,v)) &= \sum_{z \in \Gamma(w) \cap \Gamma((u,v))}  \frac{1}{\log \lvert \Gamma(z) \rvert} &
\end{flalign*}
\item Preferential Attachment (PA).
\begin{flalign*}
\text{PA}(w,(u,v)) &= \lvert \Gamma(w) \rvert \cdot \lvert \Gamma((u,v)) \rvert &
\end{flalign*}
\end{itemize}
Further, we extend the Jaccard Similarity and Adamic--Adar measures to account for a combination of the single link prediction results. We use the maximum value of the single similarity score of both end points of an edge $(u,v)$ with another node $w$, as well as the product of similarity values. We state these measure below.
\begin{itemize}
\item Jaccard Similarity.
\begin{flalign*}
\text{JS--MAX}(w,(u,v)) &= \max (JS(w,u),JS(w,v))\\
\text{JS--MUL}(w,(u,v)) &= JS(w,u) \cdot JS(w,v)
\end{flalign*}
\item Adamic--Adar.
\begin{flalign*}
\text{AA--MAX}(w,(u,v)) &= \max (AA(w,u),AA(w,v))\\
\text{AA--MUL}(w,(u,v)) &= AA(w,u) \cdot AA(w,v)
\end{flalign*}
\end{itemize}
Next, we develop two new methods for pairwise link prediction based on seeded PageRank, and use a combination of the single seeded PageRank results to compute a new measure of similarity between an edge and a node.
\subsection{Pair-seeded PageRank}
\label{sec:pairseed}
Seeded PageRank is a foundational concept in network analysis that models a flow
of information in a network to predict links and communities on a
network~\cite{Andersen-PPR-2006,Gleich-PageRank-2015}. Seeded PageRank
models information flow from the seed node to other nodes in the network via
a Markov chain, and the stationary distribution of the chain provides the scores
on the nodes. A high score on a node is a signal that the node should be
connected to the seed node.
More formally, let $\mA$ be the symmetric adjacency matrix of an undirected graph,
and let $\mP$ be the column stochastic matrix of a random walk on that graph.
Specifically, $P(i, j) = A(i, j) / \lvert \Gamma(j) \rvert$.
Let $u$ be the seed node. Then the seeded PageRank scores are entries of the 
solution vector $\vx$ to the linear system $(\mI - \alpha \mP) \vx = (1-\alpha) \ve_u$.

Here, $\ve_u$ is the vector of all zeros, except at index $u$, where $\ve_u(u) = 1$
(i.e., $\ve_u$ is the indicator vector on node $u$). 
The parameter $\alpha$ is the probability of transitioning according to the probability distribution in  $\mP$
and $(1-\alpha)$ is the probability of teleporting according to the probability distribution in $\ve_u$.
The entries of $\vx$ provide similarities between node $u$ and the other nodes and thus can be used for
standard link prediction.

In the same way seeded PageRank predicts the relevance of other nodes in the
network to a single seed node, we propose \textit{pair-seeded PageRank} to
predict the relevance of nodes to a single edge; with these similarities,
we are able to make predictions for the pairwise link prediction problem.
For a given edge $(u,v)$, pair-seeded PageRank solves the following linear system:
\[
(\mI - \alpha \mP) \vx = (1-\alpha) \ve_{u,v}.
\]
In this case, $\ve_{u,v}$ is the vector of all zeros, except at indices $u$ and
$v$, where $\ve_{u,v}(u) = \ve_{u,v}(v) = 1/2$. The solution $\vx$ can be interpreted
as the similarity of each node to the edge $(u, v)$.

We now note that pair-seeded PageRank is equivalent to the sum of single-seeded PageRank on each
of the nodes, up to a scalar multiple. This follows quickly from linearity of the
PageRank problem. To see this, let $\vx_u$ and $\vx_v$ be
the seeded PageRank solutions corresponding to nodes $u$ and $v$
respectively. Then,
\begin{align*}
(\mI - \alpha \mP) \vx_u &= (1-\alpha) \ve_{u} \\
(\mI - \alpha \mP) \vx_v &= (1-\alpha) \ve_{v} \\
\shortintertext{Adding the above two equations yields}
(\mI - \alpha \mP) (\vx_u+\vx_v) &= (1-\alpha) (\ve_{u} + \ve_{v}) \\
(\mI - \alpha \mP) (\vx_u+\vx_v) &= (1-\alpha) (2\ve_{u,v}) \\
\frac{1}{2}(\mI - \alpha \mP) (\vx_u + \vx_v) &= (1-\alpha) \ve_{u,v}\\
(\mI - \alpha \mP) \vx &= (1-\alpha) \ve_{u,v}
\end{align*}
Hence, $2 \vx = \vx_u + \vx_v$, and the pair-seeded PageRank solution is
equivalent to the summation of the single seeded PageRank equations, up to scaling.
Indeed, this is a useful and helpful observation as there are many 
systems designed to estimate large seeded PageRank values for single-seeds
by using highly scalable random walk 
methods~\cite{Lofgren-bidirectionalPR-2016}. Thus,
this technique could be used wherever a PageRank-style prediction
is already employed. 

\subsection{Triangle Reinforced PageRank (TRPR)}
\label{sec:trpr}
We now propose a PageRank-like method that uses a weighting scheme on edges
based on the number of triangles that contains each edge, which we call
\emph{Triangle Reinforced PageRank} (TRPR). For an unweighted graph, the
PageRank solution is highly affected by the degree of nodes in the network.
Here, we \textit{reinforce} the influence of triangles by giving edges participating in many
triangles a higher weight.
Figure~\ref{fig:trpr} presents a motivating example for the usefulness of reinforcing triangles.
\begin{marginfigure}
\centering
    \includegraphics[width=0.5\linewidth]{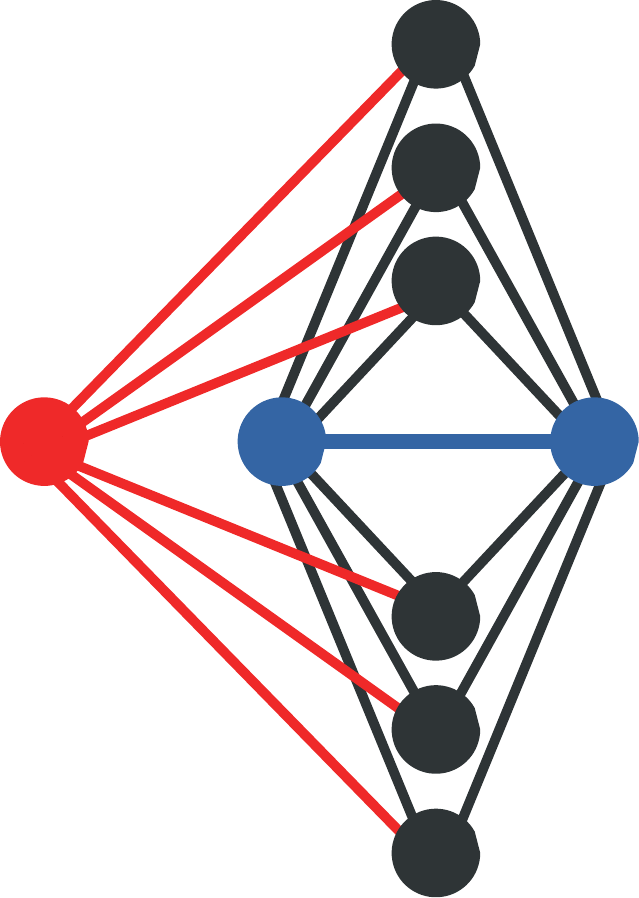}
    \caption{Motivating social network example for the TRPR algorithm. If all of the
      \textit{friends} of the blue couple know the red node, we want to
      predict that the red node must know the blue couple as well. Running
      TRPR on the above example with $\ve_{u,v}$ as the seed vector on the
      blue nodes reveals that the red node has the third highest score after
      the two blue nodes. After $10$ iterations of Algorithm~\ref{alg:trpr} with
      $\alpha = 0.85$, the output vector assigns a score of $0.120$ to the red node,
      $0.062$ to the black nodes, and $0.252$ to the blue nodes.}
    \label{fig:trpr}
\end{marginfigure}

To develop our TRPR method, we first introduce a tensor $\cmT$, that encodes all
triangles in a network:
\[
\cmT(i,j,k) =
\begin{cases}
  1 & \text{if $(i,j,k)$ is a triangle} \\
  0 & \text{otherwise}.
\end{cases}
\]
Again, in our derivation, we assume that the graph is undirected so that $\cmT$
is fully symmetric in all permutations of indices. 
A typical way to solve the PageRank linear system is the power method. With
TRPR, we modify the power method by adding a step that redistributes the weights
in the network.  Specifically, we compute the matrix $\hat{\mX} = \cmT[\vx]$,
where $\hat{\mX}(i, j) = \sum_{k}\cmT(i, j, k)\vx(k)$, which measures the
relevance of edge $(i,j)$ to the distribution of node scores in the vector
$\vx$.  We then run an iteration of the power method on a weighted adjacency
matrix $\mX = \hat{\mX} + \mA$, where the columns are re-normalized to make the
matrix column stochastic. Algorithm~\ref{alg:trpr} shows the idealized algorithm.
\begin{algorithm}[tb]
\KwIn{$\cmT,\text{adjacency matrix of undirected graph }\mA,\alpha,\ve_{u,v},\text{nb. iterations } n$}
\KwOut{$\vx$}
$\vx_0 = \ve_{u,v}$ \\
\For {$i = 1, 2, \ldots, n$}{
$\hat{\mX}^{(i)} = \cmT [\vx_{i-1}] \text{ \# i.e., $\hat{X}^{(i)}_{r,s} = \sum_{k} \cmT(r,s,k)\vx_{i-1}(k)$}$\\
$\mP_i = \text{normalize(}\hat{\mX}^{(i)} + \mA\text{)} \text{ \# column stochastic}$\\
$\vx_i = \alpha \mP_i \vx_{i-1} + (1-\alpha) \vx_0$
}
\Return{$\vx_n$}
\caption{{\bf TRPR} \label{alg:trpr}}
\end{algorithm}

\xhdr{TRPR can be implemented efficiently} 
Although TRPR involves the tensor $\cmT$, we do not need to form it explicitly, and we show an alternative derivation here. We first unwrap one iteration of TRPR. Let $\mA_i = \cmT [\vx_{i-1}] + \mA$, then, at iteration $i$, we can translate $\vx_i = \alpha \mP_i \vx_{i-1} + (1-\alpha)\vx_0$ into
\begin{align*}
\vx_i &= \alpha ((\cmT [\vx_{i-1}] + \mA) \mD_{A_i}^{-1})\vx_{i-1} + (1-\alpha)\vx_0
\end{align*}
where $\mD_{A_i}^{-1}$ is a diagonal matrix with the $i^{th}$ diagonal entry being the inverse of the sum of edge weights connected to node $i$ in $\mA_i$ (again, we assume a connected graph so these values are all non-zero). Then,
\[
\vx_i = \alpha \cmT [\vx_{i-1}] \mD_{A_i}^{-1}\vx_{i-1} + \alpha \mA \mD_{A_i}^{-1}\vx_{i-1} + (1-\alpha)\vx_0.
\]
Set $\vy_{i-1} =  \mD_{A_i}^{-1}\vx_{i-1}$. Then
\[
\vx_i = \alpha \cmT [\vx_{i-1}] \vy_{i-1} + \alpha \mA \vy_{i-1}+ (1-\alpha)\vx_0.
\]
The relevant computationally expensive pieces to compute are $\cmT[\vx_{i-1}]\vy_{i-1}$ and the entries of $\mD^{-1}_{A_i}$. Both  involve the same type of operation. Using the definition of $\cmT[\vx]$ we have that the matrix-vector product $\vz = \cmT[\vx] \vy$ has $z_i = \sum_{j} \sum_{k} \cmT(i,j,k) y(j) x(k)$. Consequently, if we have any means of \emph{iterating} over the triangles of a graph, then we can compute $\cmT[\vx] \vy$ for any pair $\vx$ and $\vy$ in a fashion akin to a sparse-matrix-vector product but in runtime proportional to the number of triangles in the graph. 

\begin{marginfigure}
    \includegraphics[width=1.1\textwidth]{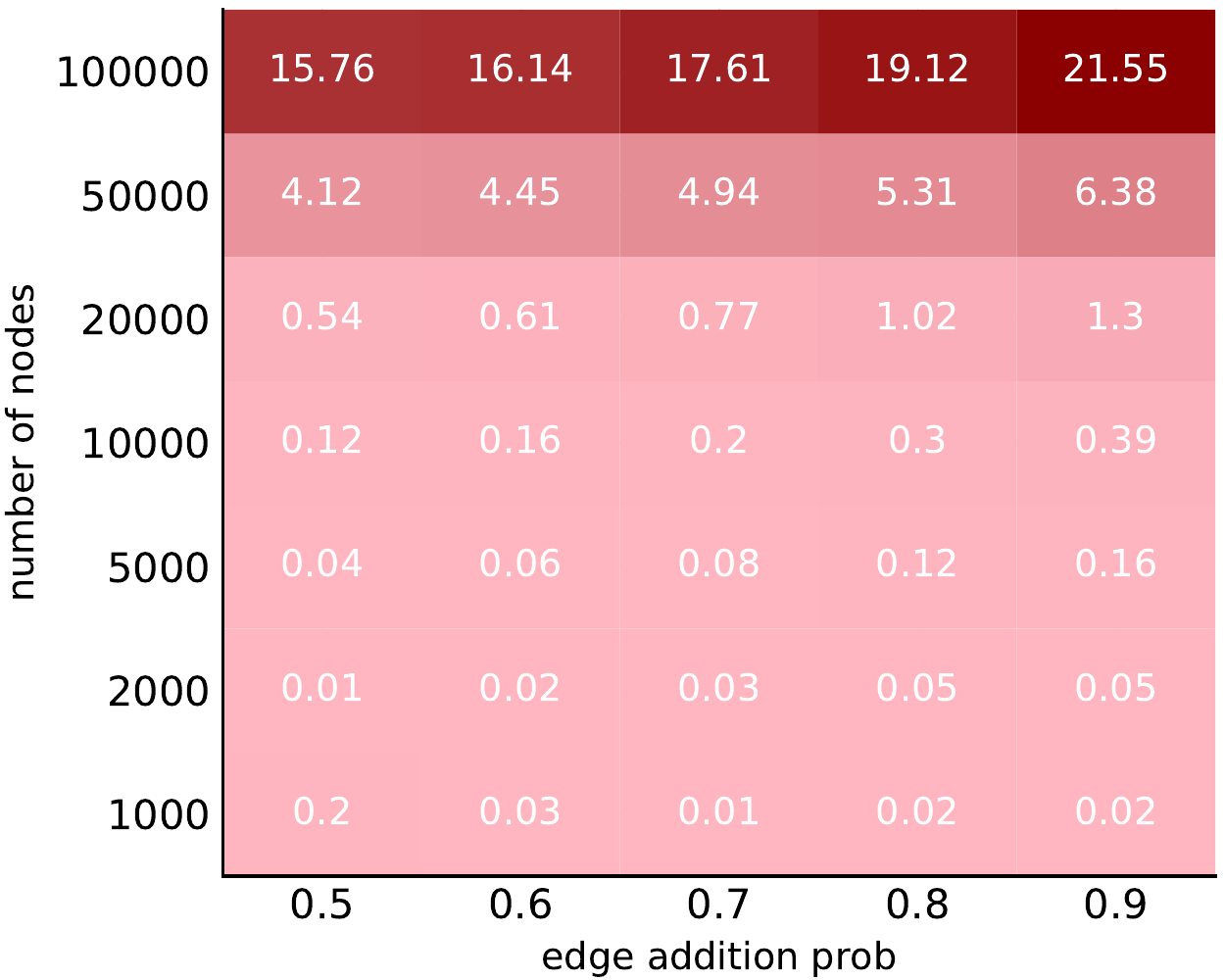}
    \caption{Time in seconds as we run TRPR for 10 iterations on generalized preferential attachement graphs as we vary the size of the network and the edge addition probability.}
    \label{fig:trpr_scale}
\end{marginfigure}

This directly enables us to compute $\cmT[\vx_{i-1}]\vy_{i-1}$. To compute the entries in $\mD^{-1}_{A_i}$, note that $\cmT[\vx]$ is a symmetric matrix because it can be written as a sum of symmetric matrices (since $\cmT$ is fully symmetric in all permutations). Thus, the row-sums of $\mA_i$ are the vertex-degrees we need to build $\mD^{-1}_{A_i}$. Let $\ve$ be the vector of all ones; these row sums are computed as $\mA_i \ve = \cmT[\vx_i] \ve + \mA \ve$. Since $\mA$ is not changing, we only need to compute the column sums of $\cmT[\vx_i] \ve$ at each iteration. Again, we can use an implicit tensor-vector-vector product operation to compute the column sums. And thus, all operations involving the tensor $\cmT$ are linear in terms of the number of triangles in the network, 
and we use a fast routine to iterate through triangles in a graph. For ease of reuse, we provide the code for TRPR at \url{https://github.com/nassarhuda/pairseed/blob/master/trpr.jl}. We experimentally validate the running time of TRPR on a preferential attachement graph while varying the size of the graph. We specifically use the generalized preferential attachment model~\cite{Avin-GPA-2017} that generalizes the classical preferential attachment model~\cite{Newman-growing-networks-2001}. In this experiment we vary the edge addition probability $p_e$, and allow the node addition probability to be $1-p_e$. Figure~\ref{fig:trpr_scale} shows the running time in seconds and empirically verifies that TRPR is a fast method when implemented efficiently and thus, is scalable to large graphs.

\xhdr{A weighted version of TRPR}
Although TRPR introduces higher weights to edges participating in many triangles by forming a new adjacency matrix $\hat{\mX} + \mA$, these weights are often dominated by the weights in the adjacency matrix $\mA$. To give a fair contribution to these edges, we introduce a scalar multiple to $\hat{\mX}$. A straightforward scalar we choose is $\gamma = \text{sum}(\mA)/\text{sum}(\hat{\mX})$. This scalar will guarantee that the sum of weights in $\mA$ and $\gamma\hat{\mX}$ are equal. We present the idealized algorithm of the weighted version of TRPR in Algorithm~\ref{alg:trprw}.

\begin{algorithm}[tb]
\KwIn{$\cmT,\text{adjacency matrix of undirected graph }\mA,\alpha,\ve_{u,v},\text{nb. iterations } n$}
\KwOut{$\vx$}
$\vx_0 = \ve_{u,v}$ \\
\For {$i = 1, 2, \ldots, n$}{
$\hat{\mX}^{(i)} = \cmT [\vx_{i-1}] \text{ \# i.e., $\hat{X}^{(i)}_{r,s} = \sum_{k} \cmT(r,s,k)\vx_{i-1}(k)$}$\\
$\gamma = \text{sum}(\mA)/\text{sum}(\hat{\mX}^{(i)})$\\
$\mP_i = \text{normalize(}\gamma\hat{\mX}^{(i)} + \mA\text{)} \text{ \# column stochastic}$\\
$\vx_i = \alpha \mP_i \vx_{i-1} + (1-\alpha) \vx_0$
}
\Return{$\vx_n$}
\caption{{\bf TRPR-Weighted} \label{alg:trprw}}
\end{algorithm}
\begin{marginfigure}
    \includegraphics[width=1.1\linewidth]{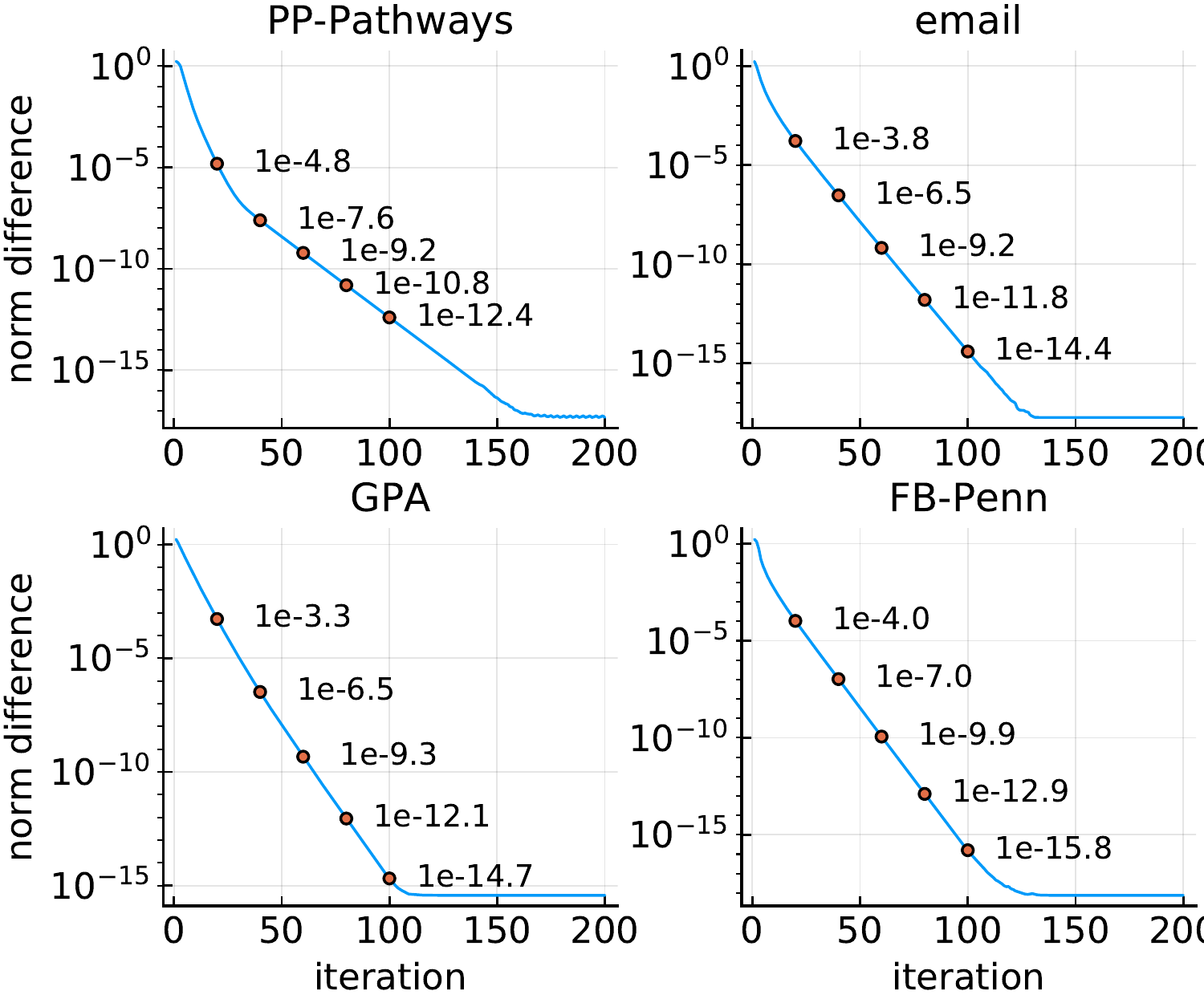}
    \caption{1-norm convergence of TRPR on 4 datasets used in the experiments section. These figures show that TRPR converges experimentally.}
    \label{fig:trpr_convergence}
\end{marginfigure}%

\begin{marginfigure}
\centering
    \includegraphics[width=1.1\linewidth]{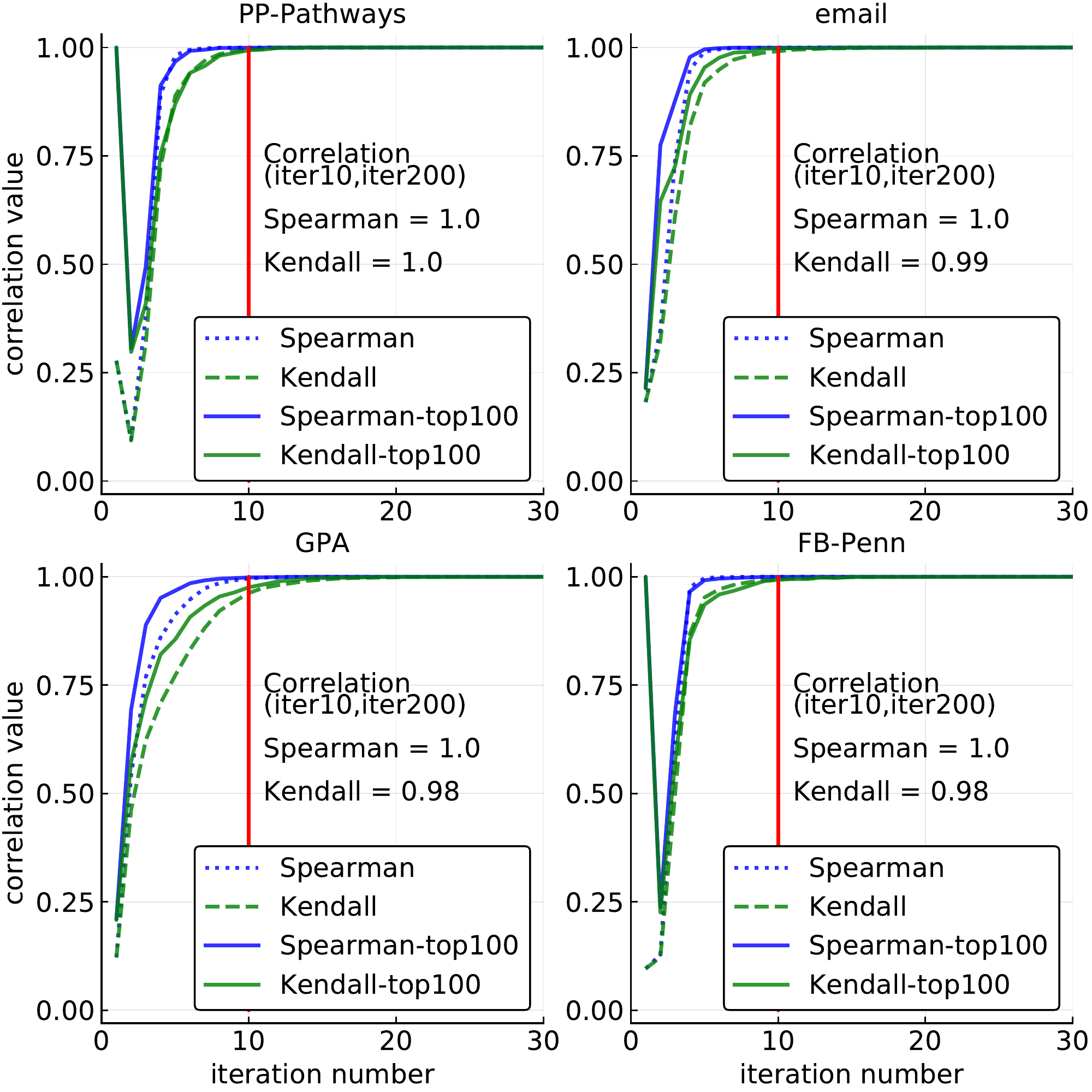}
    \caption{Spearman's rank correlation coefficient and the Kendall rank correlation coefficient between two consecutive iterates from TRPR. The solid plots show the consecutive correlation values when truncating the vectors to take the top 100 nodes, and the dashed lines compare the orderings in the full vectors. The vertical red line represents the $10^{th}$ iterate. The text in the figures is the correlation between the $10^{th}$ iterate and the $200^{th}$ iterate. These correlations support our choice of 10 iterations in the experiments involving TRPR.}
    \label{fig:trpr_order}
\end{marginfigure}

\xhdr{Convergence of TRPR}
Convergence of this type of nonlinear system of equations is theoretically delicate with bounds that are often insufficient for practice~\cite{Benson-Spacey-2016}. Empirically, we observe that the iterations converge. However, absent a robust theory, this method is only run for a small and fixed number of iterations (10). This will produce a unique deterministic and reproducible set of scores that locally capture the influence of both the graph and the reinforced triangles. In Figure~\ref{fig:trpr_convergence}, we show the 1-norm difference decay from two consecutive iterates from TRPR on 4 datasets used in the experiments section. Figure~\ref{fig:trpr_convergence} shows that the method converges experimentally. Even though the norm convergence seems to happen after around 100 iterations in these datasets, the ordering of nodes in these vectors does not change much after a few iterations. We run another experiment to study the ordering of the nodes from every iteration and notice that the order does not change much after just a few iterations. In Figure~\ref{fig:trpr_order}, we show the Spearman's rank correlation coefficient and the Kendall rank correlation coefficient between two consecutive iterates from TRPR on the same four graphs used in Figure~\ref{fig:trpr_convergence}. We notice that after a few iterations (10) the orderings of the vectors no longer change, and especially the order of the top 100 nodes does not change (solid lines in the plot in Figure~\ref{fig:trpr_order}).

\subsection{Extensions of single-seeded PageRank}
\label{sec:pr-extension}
We also use the single seeded PageRank solution of each endpoint of the edge we are interested in predicting links to and produce two more metrics for relating an edge to a node. Denote $\vx_u$, and $\vx_v$ to be the seeded PageRank solutions for nodes $u$ and $v$ respectively. Then, we define \textit{MAX} and \textit{MUL} as follows.
\begin{flalign*}
\text{MAX}(u,v) &= \max (\vx_u,\vx_v) \text{ (element-wise maximum)}\\
\text{MUL}(u,v) &= \vx_u \odot \vx_v \text{ (element-wise multiplication)}
\end{flalign*}

\hide{
\section{Methods}
To solve pairwise link prediction, we propose a number of methods. First, we extend three  baseline local metrics from the node-node similarity scenario to node-edge similarity. Second, we propose three diffusion type methods akin to PageRank. We describe our proposed methods below.
\subsection{Local measures extensions}
\label{sec:local}
As in the link prediction scenario, local link prediction methods for pairwise link prediction is dependent on the neighborhoods of the nodes and edges. First, we define the neighborhood of an edge $e = (u,v)$, to be the set $\Gamma(e) = \{w|w \text{ forms a triangle with }e\}$, or alternatively $\Gamma(e) = \Gamma(u) \cap \Gamma(v)$, where $\Gamma(u)$ is the set of nodes adjacent to node $u$. Social networks often evolve in a way dependent on the neighborhoods of the nodes, and thus, such extensions will prove to be useful specially in cases of predicting links on social networks.

\begin{itemize}
\item Jaccard Similarity.
\begin{flalign*}
JS(w,(u,v)) &= \frac{|\Gamma(w) \cap \Gamma((u,v))|}{|\Gamma(w) \cup \Gamma((u,v))} &
\end{flalign*}
\item Adamic Adar.
\begin{flalign*}
AA(w,(u,v)) &= \sum\limits_i \frac{1}{\text{log}|\Gamma(i)|}, i \in \Gamma(w) \cap \Gamma((u,v)) &
\end{flalign*}
\item Preferential Attachment.
\begin{flalign*}
PA(w,(u,v)) &= |\Gamma(w)||\Gamma((u,v))| &
\end{flalign*}
\end{itemize}

\subsection{Pair-seeded PageRank}
\label{sec:pairseed}
Seeded PageRank is a foundational concept in network analysis that models a flow of information in a network to predict links and communities on a network. The main premise of seeded PageRank is that it models information/score transport from the seed node to other nodes in the network. When a different node receives a high score, this is signal to deduce that this node has a high likelihood to be connected to the seed node.

In the same way seeded PageRank predicts the relevance of other nodes in the network to a single seed node, we propose \textit{pair-seeded PageRank} to predict the relevance of nodes to a single edge.

Let $\mP$ be the column stochastic matrix of the adjaceny matrix corresponding to a graph $G_A$. And let $u$ be the node of interest, in which we are interested to predict links to, we call $u$ the seed node, seeded PageRank solves the following linear system.
\[
(\mI - \alpha \mP) \vx = (1-\alpha) \ve_u
\]
where $\ve_u$ is the vector of all zeros, except at index $u$, $\ve_u(u) = 1$. For a given edge $(u,v)$, pair-seeded PageRank solves the following linear system.
\[
(\mI - \alpha \mP) \vx = (1-\alpha) \ve_{u,v}
\]
where $\ve_{u,v}$ is the vector of all zeros, except at indices $u$ and $v$, $\ve_{u,v}(u) = 1/2$, and $\ve_{u,v}(v) = 1/2$. Pair-seeded PageRank is equivalent to the added soution of single seeded PageRank on each of the nodes, up to a scalar multiple. Let $\vx_u$ and $\vx_v$ be the seeded PageRank solutions corresponding to nodes $u$ and $v$ respectively. Then,
\begin{align*}
(\mI - \alpha \mP) \vx_u &= (1-\alpha) \ve_{u} \\
(\mI - \alpha \mP) \vx_v &= (1-\alpha) \ve_{v} \\
\shortintertext{Adding the above two equations yields}
(\mI - \alpha \mP) (\vx_u+\vx_v) &= (1-\alpha) (\ve_{u} + \ve_{v}) \\
(\mI - \alpha \mP) (\vx_u+\vx_v) &= (1-\alpha) (2\ve_{u,v}) \\
\frac{1}{2}(\mI - \alpha \mP) (\vx_u + \vx_v) &= (1-\alpha) \ve_{u,v}\\
(\mI - \alpha \mP) \vx &= (1-\alpha) \ve_{u,v}
\end{align*}
Hence, $2 \vx = \vx_u + \vx_v$, and the pair-seeded PageRank solution is equivalent to the summation of the single seeded PageRank equaitons, up to a scalar.

\subsection{Triangle Reinforced PageRank (TRPR)}
\label{sec:trpr}
Triangle Reinforced PageRank (TRPR) is our method to impose higher weights on important edges that participate in many triangles in a graph. For an unweighted graph, the PageRank solution is highly determined by the degree of nodes in the network, and here, we \textit{reinforce} triangles by giving edges participating in many triangles a higher weight. To do so, we first introduce a tensor $\cmT$, that encodes all triangles in a network. $\cmT(i,j,k) = 1$ if $(i,j,k)$ is a triangle in the graph. Then, we modify the power method which is the standard way to solve the PageRank linear system by adding one step that redistributes the weights in the network. Specifically, the $(i,j)$ entry in the matrix $\mX = \cmT \vx$ is the score of edge $(i,j)$ relevant to the distribution of node scores in the vector $\vx$ (line 5 in Algorithm~\ref{alg:trpr}). We present a motivating example of the TRPR algorithm in Figure~\ref{fig:trpr}.
\begin{algorithm}
\KwIn{$\cmT,\mA,\alpha,\ve_{u,v},iters$}
\KwOut{$\vx$}
$\vx_0 = \ve_{u,v}$ \\
\While {iterations not finished}{
$\hat{\mX}_i = \cmT \vx_{i-1}$\\
$\mX_i = \text{normalize(}\hat{\mX}_i + \mA\text{)} \text{ \# column stochastic}$\\
$\vx_i = \alpha \mX_i \vx_{i-1} + (1-\alpha) \vx_0$
}
\caption{{\bf TRPR} \label{alg:trpr}}
\end{algorithm}

\begin{figure}
	\begin{minipage}{0.3\linewidth}
		\includegraphics[width=\linewidth]{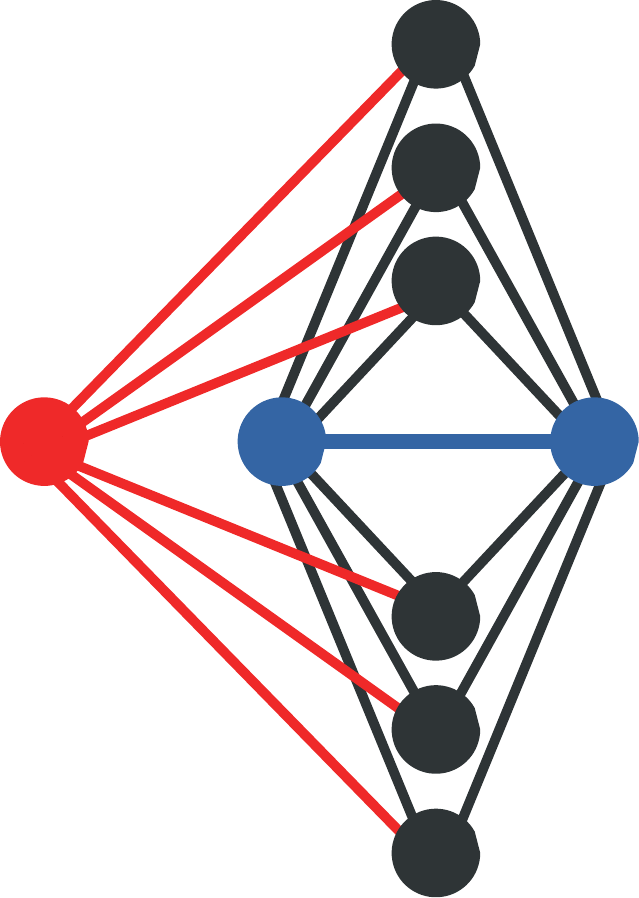}
	\end{minipage}
\begin{minipage}{0.6\linewidth}
	\caption{A motivating example of the TRPR algorithm. If all the \textit{friends} of the blue couple know the red node, we want to predict that the red node must know the blue couple as well. Running TRPR on the above example, with $\ve_{u,v}$ being a seed vector on the blue nodes reveals the red node being the third highest score after the two blue nodes. For $10$ iterations, and $\alpha = 0.8$, the ouput vector evaluates the red node to have a score $0.102$, the black nodes having a score $0.063$, and the blue nodes having a score $0.257$.}
	\label{fig:trpr}
\end{minipage}
\end{figure}

[XXX] Show equivalence to David's notation.

%
\subsection{Collapsed network}
Another approach to solve the problem of pairwise link prediction is to \textit{transform} the problem. In this method, we transform the graph to a new graph where each pairwise connection represents a triangle in the old graph. Then, we can perform seeded PageRank for link prediction. The key idea is, for every triangle, treat two of the nodes as one new node (i.e. collapse them to a new node in the new graph), and form a edge between the collapsed node and the third node for the corresponding triangle. Refer to figure~\ref{fig:collapse} for an illustration on how we transorm a small triangle graph. All 6 enumerations of the triangle are encoded via edges and every node with two colors encodes two previous nodes. Encoding all possible enumerations of the triangle allows for an easeful extension to directed graphs. 

This problem transformation may be a natural one, but our experimental evidence has shown that using \textit{only} triangles to predict triangles is not useful. Also, this method is not easily applicable to any graph as it can grow the original graph of $n$ nodes to a new graph of size $n^2$. Due to this method's computational expensiveness and lack of good performance, we limit its use to the synthetic graphs experiments and demonstrate that it is only beneficial to use it for triangle dense graphs, i.e. graphs where the ratio of the number of triangles multiplied by 3 to the number of wedges in the graph is high. In the experiments section, we verify this by applying the network collapsion method on a clique graph (which has a triangle density = 1).
\begin{figure}[h]
  \centering
  \includegraphics[width=0.5\linewidth]{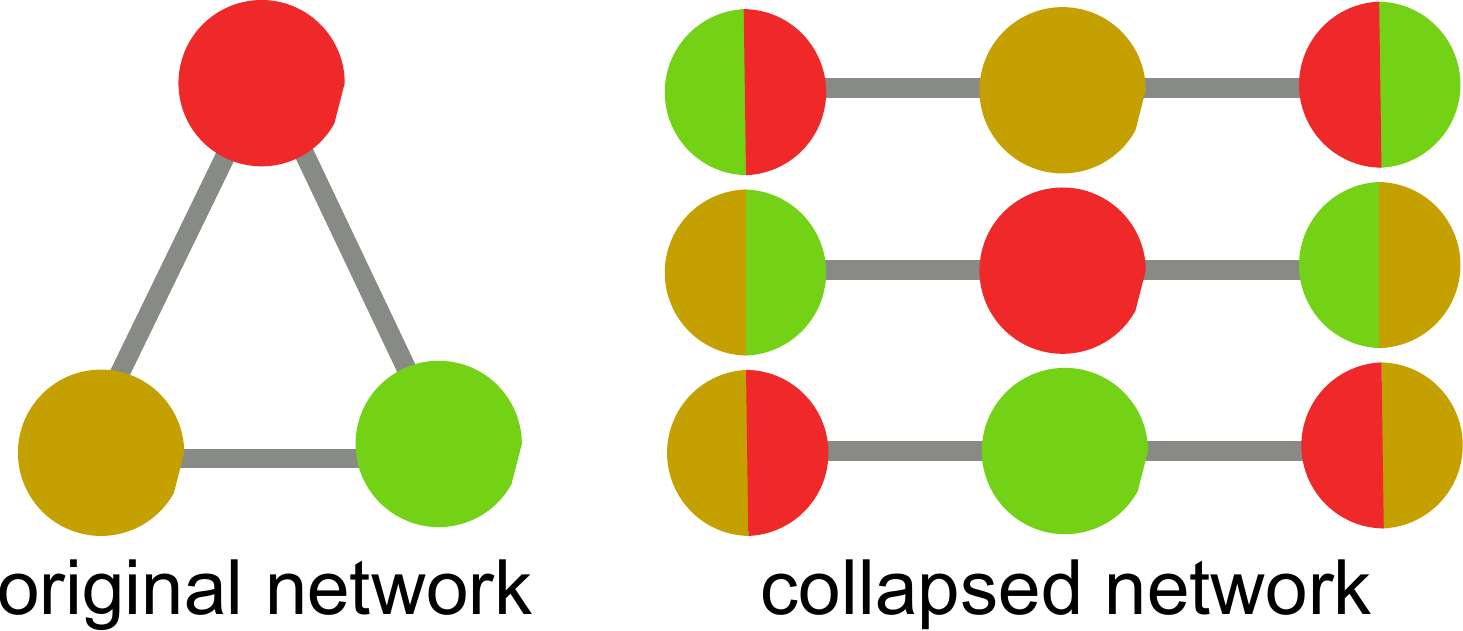}
  \caption{An illustration of the network transformation of collapsing a triangle graph to a new graph where each edge in the new graph represents a triangle in the old graph.}
  \label{fig:collapse}
\end{figure}
}

%% file: 004-evaluations.tex
\section{Experimental Setup}

We now perform a series of experiments on synthetic as well as real-world graphs
from a variety of disciplines, including online social networks, communication
networks, and biological interaction networks. We also include experiments for
static networks as well as a temporal network. For evaluation, we use the Success Probability (SP) measure, 
which we define for one experiment as follows:

\[
\text{SP}((u,v),k) = \begin{cases}
  1 & \text{if at least one ground truth node $w$ appears in the top $k$} \\
     & \text{predictions for edge $(u,v)$}\\
  0 & \text{otherwise}.
\end{cases}
\]

Note that the top $k$ predictions is the set of $k$ nodes that are not connected to either end point of the seed edge 
with the highest scores.
For our experiments, we will have training data and validation data, and the ground truth nodes that should be connected to an edge from the training data can be deduced from the validation data. For each training dataset, we run $500$ random experiments, where we try to predict links to $500$ randomly chosen edges (we call them \textit{seed edges}). For each experiment, an SP value ($0$ or $1$) is computed, and the overall score is the mean value over all the experiments.
The main choice for this measure in contrast to the area under ROC curve (AUC score) measure for instance, is the small number of nodes that we often want to recover. For a given edge $(u,v)$ in the training data, the validation data must have the edges $(u,w)$ and $(v,w)$ for $w$ to be considered a correct ground truth node for recovery. In subsequent sections, we will see that the number of nodes that satisfy this property in the validation data is often small ($1$ in most instances), and thus a measure such as the AUC score does not fully capture the performance of our methods.

\subsection{Leave One Edge's Triangles Out (LOETO)}
The LOETO experiments are akin to the leave-$p$-out cross validation metric, in the sense that we will use $p$ edges as a validation set and the remaining edges of the network as a training set; 
here, $p = (2$ $\times$ number of nodes that form a triangle with a randomly chosen edge$)$. An experimental trial in this setting is designed as follows. Randomly pick an edge in the graph (call it the seed edge) and find all the wedges (path of length 2) that form a triangle with this edge. Next, drop all these wedges and place them in the validation set. Figure~\ref{fig:LOETO} visualizes this experiment. The graph used will be the one in panel B of Figure~\ref{fig:LOETO} (the grey dashed edges no longer appear in the network and the goal is to recover the connections with the green nodes in the graph). We then use the pairwise link prediction methods on the seed edge, which
produces an ordering on the nodes, and given this ordering, we compute the success probability. Since this method leaves a big portion of the graph in the training data, we compute its Success Probability with top $k=5$.

\subsection{Hold-out cross validation}
The hold-out cross validation method that keeps a certain percentage of the data
as training set and the remaining set as validation is a standard way of
evaluating the classical link prediction problem. In this setup, for a given
network, we remove $30\%$ of the edges and label them as validation data, and
use the remaing $70\%$ as training data to make predictions.  Then, for random
edges in the training data (seed edges), we use the pairwise link prediction
methods to predict which nodes will form triangles with each edge that is
selected. Again, for a given edge, each method produces a similarity score on
all nodes, and we use the ordering of the nodes induced by the scores to calculate the
Success Probability with top $k$ values $= 5,25$.

\hide{\begin{figure}
  \begin{minipage}{0.6\linewidth}
 \includegraphics[width=\linewidth]{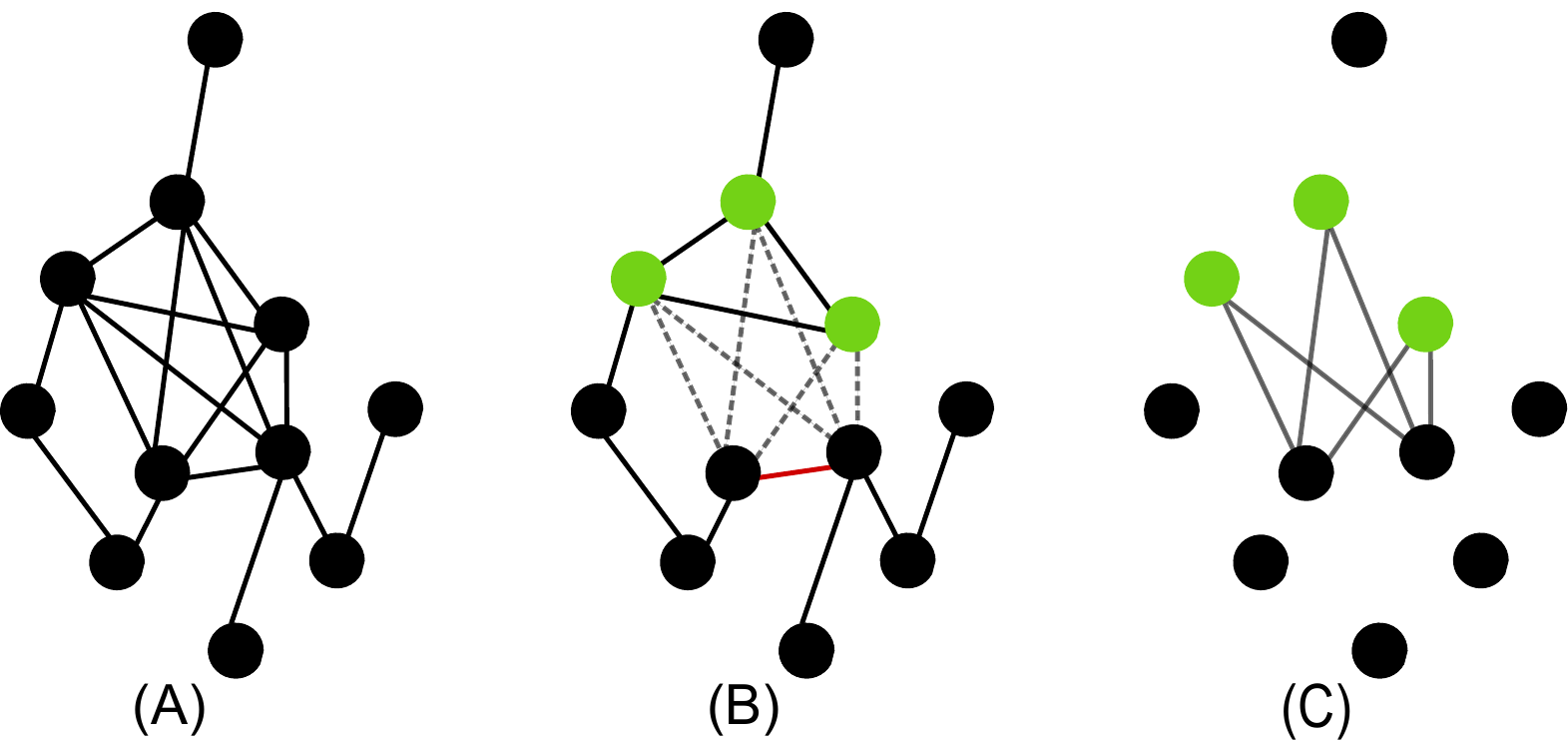}
   \end{minipage}
  \begin{minipage}{0.35\linewidth}
    \caption{Illustration of the Leave One Edge's Triangles Out (LOETO) experiment. For a given graph (subfigure A), randomly pick an edge (red edge in subfigure B), and remove all edges that form a triangle with it (dashed grey lines in subfigure B). Run all our methods on this new graph. The nodes to predict are the green nodes.}
    \label{fig:LOETO}
  \end{minipage}
\end{figure}}
\begin{marginfigure}
 \includegraphics[width=\linewidth]{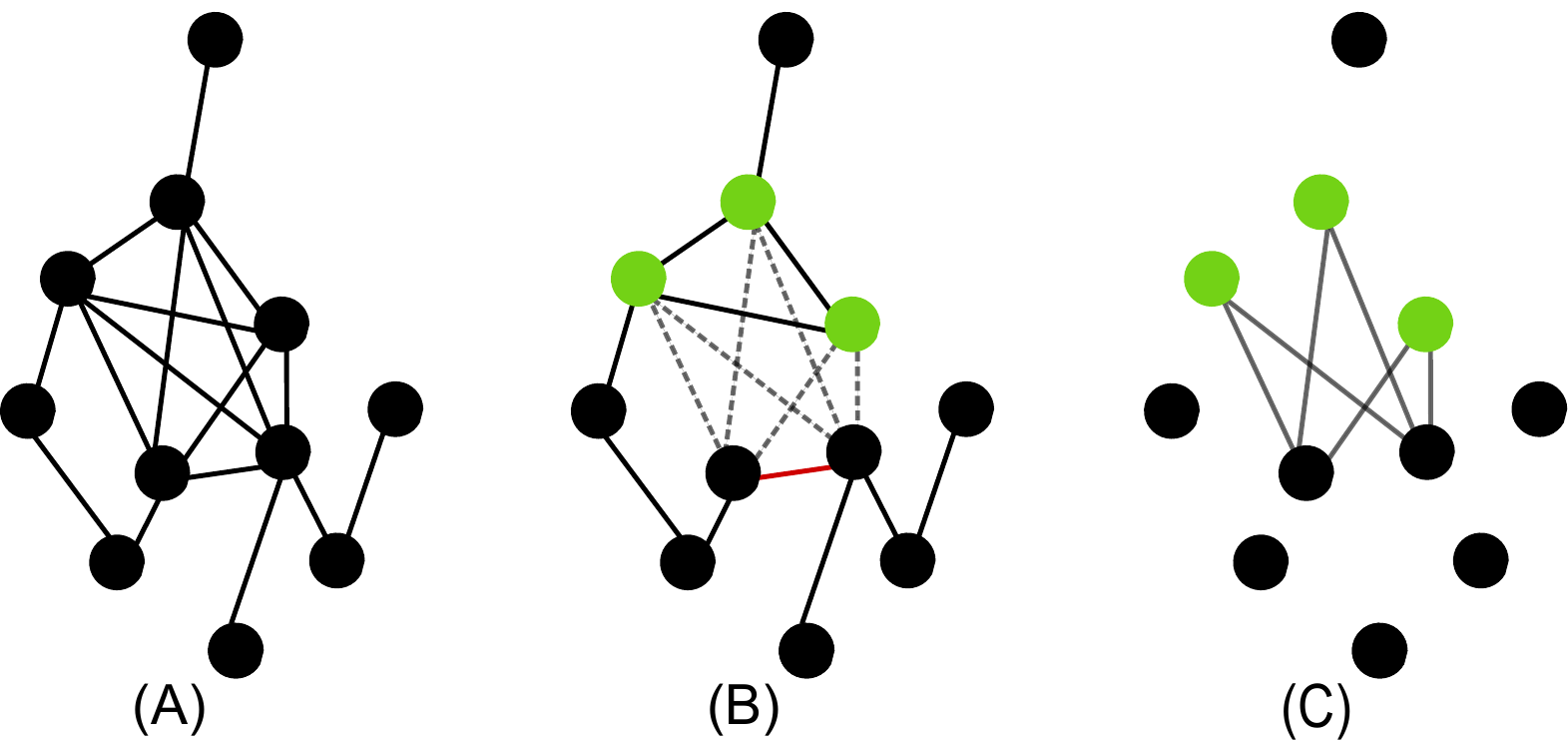}
    \caption{Illustration of the Leave One Edge's Triangles Out (LOETO) experiment. For a given graph (subfigure A), randomly pick an edge (red edge in subfigure B) and remove all edges that form a triangle with it (dashed gray lines in subfigure B). Run all our methods on this new graph. The nodes to predict are the green nodes.}
    \label{fig:LOETO}
\end{marginfigure}
We also perform a similar experiment on temporal networks with timestamps
on the edge arrivals. In this scenario, the dropped $30\%$ edges are not
chosen at random. Instead, we split the data into training and test sets
based on the time---the first $70\%$ of the edges to appear in time
are the training data and the remaining $30\%$ are the test data.

In this set of experiments, we perform one more processing step to guarantee
that the network we will use for training is connected. If the network is
disconnected, we extract the largest connected component.

\subsection{Summary of methods and parameter settings}

Finally, we summarize all of the methods that we use for pairwise link prediction.

\begin{itemize}
\item Pairseed: This is our method described in Section~\ref{sec:pairseed}. We
  use the implementation from \texttt{MatrixNetworks.jl}~\cite{MatrixNetworks}
  with $\alpha = 0.85$.  This implementation solves the linear system until
  convergence to machine precision.
\item TRPR: This is our method described in Section~\ref{sec:trpr}.
  We use $\alpha = 0.85$ and number of iterations $n = 10$.
\item TRPRW: This is the modified weighted version of the TRPR algorithm described in Section~\ref{sec:trpr} as well. We use $\alpha = 0.85$ and number of iterations $n = 10$.
\item MUL, MAX: These are the methods from Section~\ref{sec:pr-extension} that extend the single-seeded PageRank solutions. We use
  the same implementation used by Pairseed, with $\alpha=0.85$.
\item AA, PA, JS: For a seed edge, we compute the generalized Adamic-Adar,
  Preferential Attachment, and Jaccard similarity scores, respectively (as
  presented in Section~\ref{sec:local}) between the seed edge and all remaining nodes in
  the graph.
\item AA--MUL, AA--MAX, JS--MUL, JS--MAX: These are the methods from Section~\ref{sec:local}, 
and they use the single node similarity from both endpoints of a seed edge to compute a new measure of similarity.
\end{itemize}

\hide{

%

\section{Experimental Design and Methods Parameter settings}
To effectively evaluate our pairwise link prediction methods, we introduce two main experiments. Also, in the experiments section, we use a temporal network to perform a time-stamped pairseeded link prediction experiment, and present a study that shows how to utilize pairwise link prediciton techniques to perform the standard link prediction task. All these experiments fall under one of the categroies below
\subsection{Wedges experiments}
This type of experiments is a proof-of-concept, and is akin to the leave-p-out cross validation metric, in the sense that we will drop a select number of edges from the network and use them as a validation set. One random experiment under this setting is designed as follows. For a given graph, randomly pick an edge in this graph (call it the seed edge), and find all the triangles it belongs to. Then, drop one side of edges from all triangles, hence transforming all the triangles that the seed edge participates in to wedges. Then, we perform our proposed methods on the seed edge. Note that initially, we intended to remove both edges of the triangle (i.e. transforming all triangles that the seed edge participates in to edges), but this results in keeping very little higher order information around the edge and resulted in producing uniform results that weren't very meaningful in most cases. Hence, we resorted to keeping some partial information in the network.
\subsection{80-20 experiments}
This is akin to the hold-out cross validation method, and is a classical way of evaluating the standard link prediction problem. Here, for a given network, we drop $20\%$ of the edges and label them as testing data, and use the remaing $80\%$ to make predictions. Then, for random edges in the training data (call them seed edges), perform the methods proposed. To evaluate this method, we consider the verification set to consist of all nodes that formed a triangle with the seed edge, but, at least one of the edges has been removed from the training data. The main reason we consider the setting of at least one edge being removed from the training data and not both is simply because the training data is generated by a \textit{random} removal of some edges and it is unrealistic to consistently find wedges in the testing data. Nevertheless, we consider this setting in a temporal network case study in the experiments section.
\subsection{Methods Summary and paremeter settings when necessary}
\begin{itemize}
\item Pairseed: This is our method in~\ref{sec:pairseed}. We use the implementation from~\cite{MatrixNetworks} and use $\alpha = 0.8$. This implementation solves the linear system until convergence to machine precision.
\item SS: For contrasting purposes, we present the results of single seeded PageRank on one of the end points of the seed edge. We use the same implementation as above, with $\alpha=0.8$
\item TRPR: This is our method in~\ref{sec:trpr}. We use $\alpha = 0.8$, and maximum number of iterations = 10.
\item AA, PA, JS: For a seed edge, we compute the Adamic-Adar, Preferential Attachement, and Jaccard similarity scores, respectively, as presented in~\ref{sec:local} for that seed edge and all the remaining nodes in the graph.
\end{itemize}
}

%% file: 005-results.tex
\section{Pairwise link prediction results}
\label{sec:results}
In all of the results in
this section, we report the success probability from our predictions over $500$ random experiments. We use seven real-world graphs from different disciplines in this section and give a summary of their statistics in Table~\ref{tab:datasets}. We also use a synthetic graph generated from the generalized preferential attachment model (GPA)~\cite{Avin-GPA-2017}. 

\xhdr{Synthetic graph} Generalized Preferential Attachment
(GPA)~\cite{Newman-growing-networks-2001} is a synthetic graph generation model
that generalizes the classical prefernetial attachment model to allow for the
addition of new components at each step of the algorithm. For our experiments, we
generate a graph with $5000$ nodes and allow the event of node addition with
probability $1/2$, and we allow the event of edge addition with probability
$1/2$. The starting graph structure is a clique of size $5$. At each step of the
graph generation process, an edge or node is added by attaching proportionally
to the degrees of the existing nodes.

\xhdr{Real world graphs} We use various real world graphs to test our methods and provide statistics about them in Table~\ref{tab:datasets}. \textit{Penn94} and \textit{Caltech36} are online social networks from
the \textit{Facebook100} collection of datasets~\cite{Traud-facebook100-2011}. These two datasets are the biggest and smallest networks in terms of number of nodes respectively from this collection. \textit{Ch-Ch-Miner} is a biological network of drug (chemical) interactions~\cite{Wishart-drugbank-2017,biosnapnets}. \textit{P-P-Pathways} is a biological network of  physical interactions between proteins in humans~\cite{Agrawal-PPI-2018}. \textit{email} is an email communication network~\cite{Guimer-selfsimilar-2003}. 
Finally, \textit{CollegeMsg}~\cite{Panzarasa-college-2009} and \textit{email-EU}~\cite{Panzarasa-college-2009} 
are temporal networks representing private messages (CollegeMsg) or emails (email-EU) between users in a network.
  
  \begin{table}[t]
  \caption{Statistics of the real-world datasets used in this paper.}\vspace{-0.5\baselineskip}
  \label{tab:datasets}
  \begin{tabularx}{\linewidth}{lXXXX}
    \toprule
    Network name &  nodes & edges & triangles &type \\
    \midrule
    Penn94 &  41536 & 1362220 & 7207796&Social \\
    Caltech36 &  762 & 16651 &119562 &Social \\
    Ch-Ch-Miner &  1510 & 48512 &568466 &Biology \\
    P-P-Pathways &  21521 & 338624 &2394642 &Biology\\
    email &  1133 & 5451 &5343& Communication\\
    CollegeMsg &  1899 & 13838 & 14319&Temporal \\
    Email-EU &  1005 & 32128 &105461 &Temporal \\
    \bottomrule
  \end{tabularx}
\end{table}
\begin{figure}
\begin{fullwidth}
\begin{minipage}[t]{\dimexpr(\textwidth)}
  \centering\raisebox{\dimexpr \topskip-\height}{%
  \includegraphics[width=\textwidth]{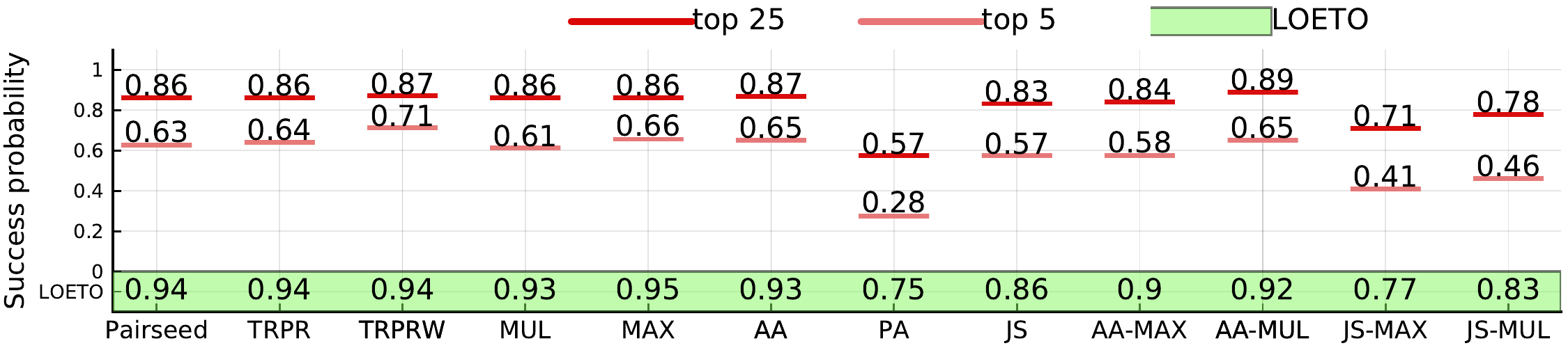}}
\end{minipage}
\begin{minipage}[t]{\dimexpr(0.9\marginparwidth)}
Ch-Ch-Miner (Bio)\\
nodes = 1510\\
edges = 48512\\
triangles = 569466\\
median(nodes to predict) = 3\\
\end{minipage}\\
\begin{minipage}[t]{\dimexpr(\textwidth)}
  \centering\raisebox{\dimexpr \topskip-\height}{%
  \includegraphics[width=\textwidth]{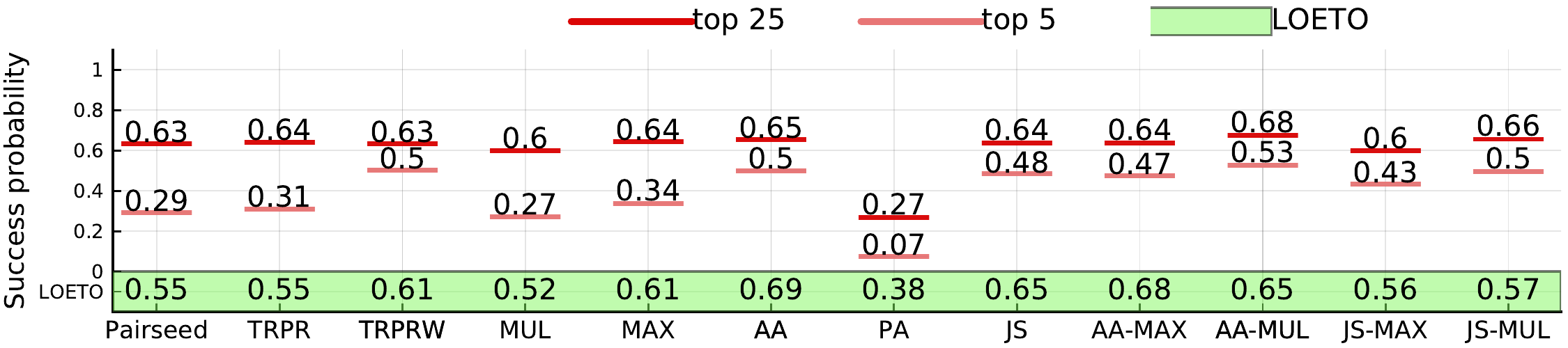}}
\end{minipage}
\begin{minipage}[t]{\dimexpr(0.9\marginparwidth)}
P-P-Pathways (Bio)\\
nodes = 21521\\
edges = 338624\\
triangles = 2394642\\
median(nodes to predict) = 2\\
\end{minipage}\\
\end{fullwidth}
\caption{Success probability results for the two biological datasets. In both datasets, we notice that TRPRW outperforms the remaining diffusion type methods and performs best on the top $k$ predictions metric on the Ch-Ch-Miner dataset. Another method that stands out in these two datasets is AA-MUL which is the best method in terms of top $k$ predictions in the P-P-Pathways dataset, with TRPRW performing worse than AA-MUL by around $5\%$ on the top $k$ measures.}
\label{fig:results_bio}
\end{figure}

\begin{figure}
\begin{fullwidth}
\begin{minipage}[t]{\dimexpr(\textwidth)}
  \centering\raisebox{\dimexpr \topskip-\height}{%
  \includegraphics[width=\textwidth]{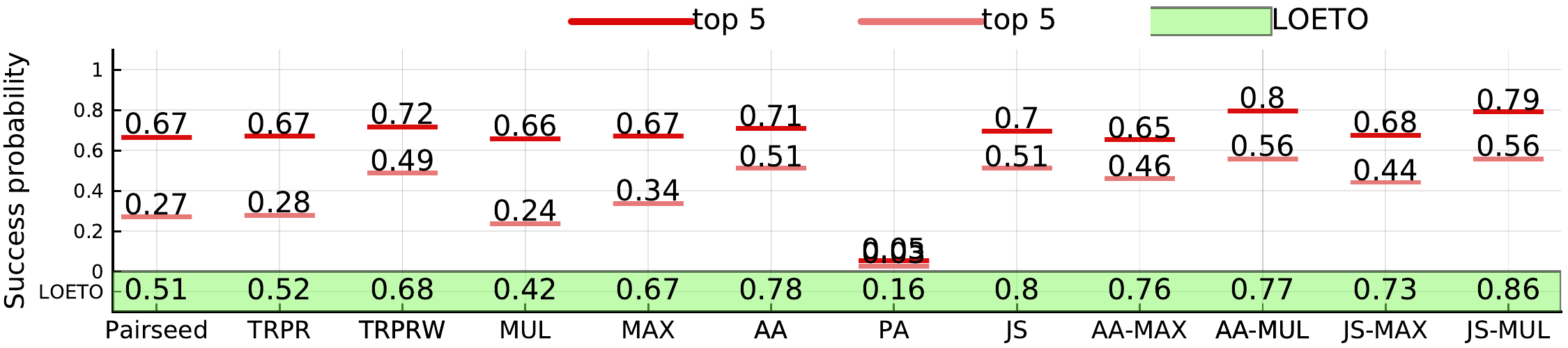}}
\end{minipage}
\begin{minipage}[t]{\dimexpr(0.9\marginparwidth)}
Facebook - Penn94 (social)\\
nodes = 41536\\
edges = 1362220\\
triangles = 7207796\\
median(nodes to predict) = 2\\
\end{minipage}\\
\begin{minipage}[t]{\dimexpr(\textwidth)}
  \centering\raisebox{\dimexpr \topskip-\height}{%
  \includegraphics[width=\textwidth]{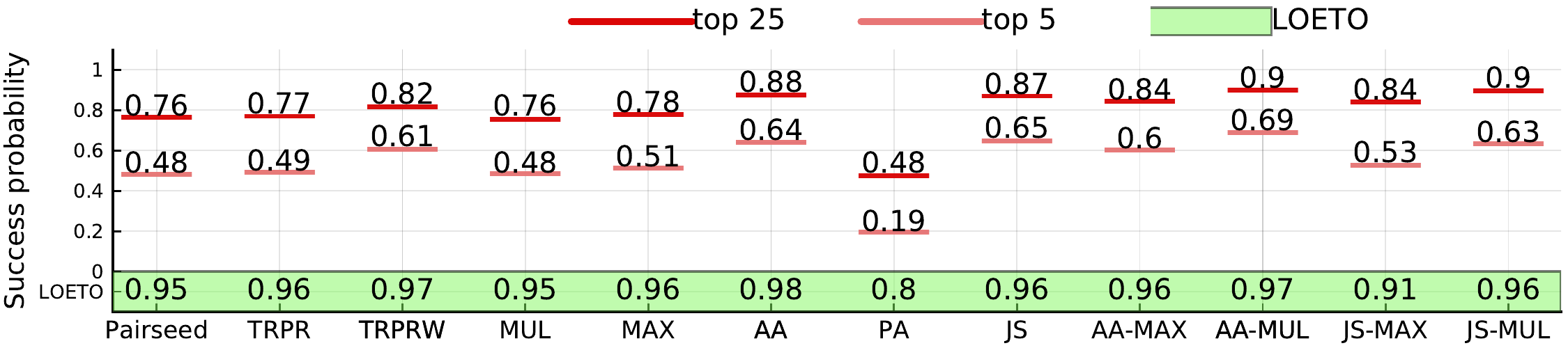}}
\end{minipage}
\begin{minipage}[t]{\dimexpr(0.9\marginparwidth)}
Facebook - Caltech36 (social)\\
nodes = 762\\
edges = 16651\\
triangles = 119562\\
median(nodes to predict) = 2\\
\end{minipage}\\
\end{fullwidth}
\caption{Success probability results for the two social networks datasets. In both datasets, we notice that local methods generally outperform diffusion type methods. This is mainly due to how social networks grow and the influence of neighbors of nodes for making new connections. Here too, TRPRW outperforms other diffusion type methods and produces comparable results to the best local methods on the top 25 and LOETO measures.}
\label{fig:results_fb}
\end{figure}

\begin{figure}
\begin{fullwidth}
\begin{minipage}[t]{\dimexpr(\textwidth)}
  \centering\raisebox{\dimexpr \topskip-\height}{%
  \includegraphics[width=\textwidth]{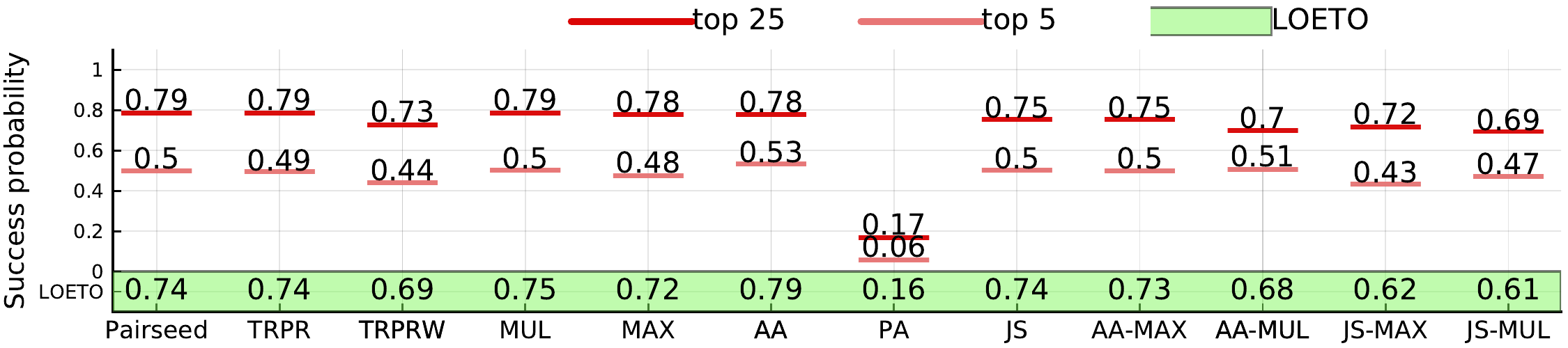}}
\end{minipage}
\begin{minipage}[t]{\dimexpr(0.9\marginparwidth)}
email (communication)\\
nodes = 1133\\
edges = 5451\\
triangles = 5343\\
median(nodes to predict) = 1\\
\end{minipage}\\
\begin{minipage}[t]{\dimexpr(\textwidth)}
  \centering\raisebox{\dimexpr \topskip-\height}{%
  \includegraphics[width=\textwidth]{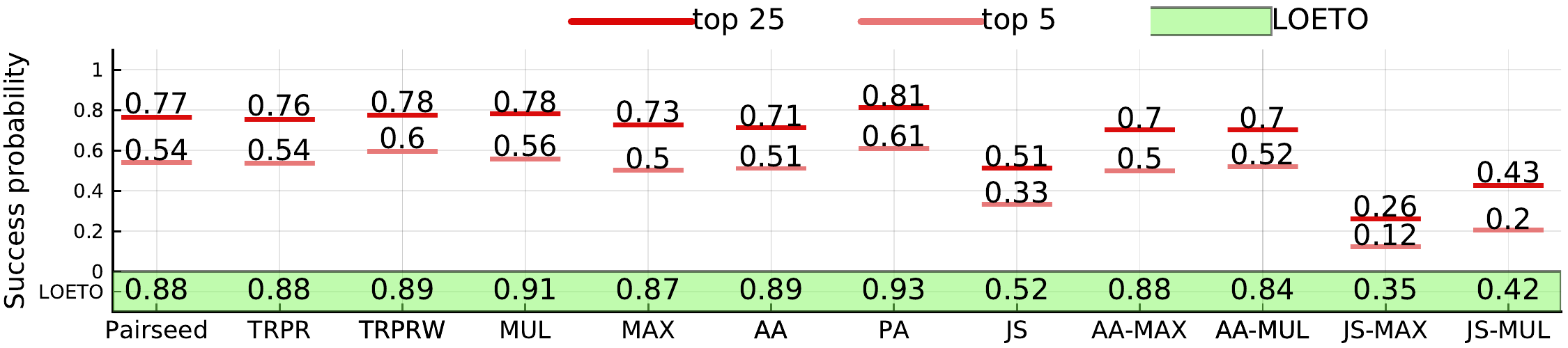}}
\end{minipage}
\begin{minipage}[t]{\dimexpr(0.9\marginparwidth)}
GPA graph (synthetic)\\
p(node addition) = 0.5\\
p(edge addition) = 0.5\\
triangles = 5897\\
median(nodes to predict) = 1\\
\end{minipage}\\
\end{fullwidth}
\caption{Success probability results for the two networks, email and an instance of a GPA graph. We group these two graphs together because they have a very small number of triangles compared to the other networks. In these datasets TRPRW does not contribute an improvement over the other diffusion type methods. In the email network, TRPR performs best in the top $k$ metric, and TRPRW performs best after the PA method on the GPA graph.}
\label{fig:results_pa}
\end{figure}

\begin{figure}
\begin{fullwidth}
\begin{minipage}[t]{\dimexpr(\textwidth)}
  \centering\raisebox{\dimexpr \topskip-\height}{%
  \includegraphics[width=\textwidth]{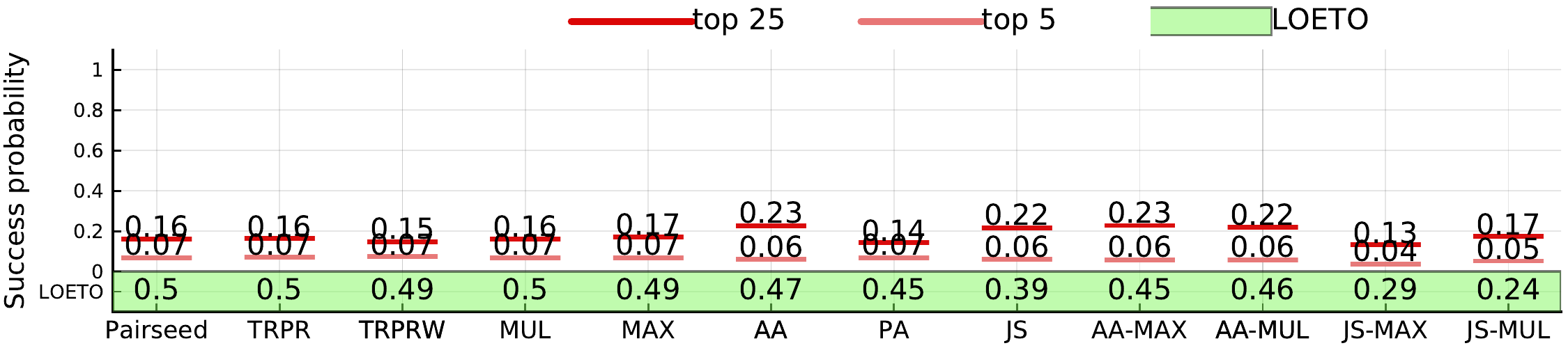}}
\end{minipage}
\begin{minipage}[t]{\dimexpr(0.9\marginparwidth)}
CollegeMsg (temporal)\\
nodes = 1899\\
edges = 13838\\
triangles = 14319\\
median(nodes to predict) = 1\\
\end{minipage}\\
\begin{minipage}[t]{\dimexpr(\textwidth)}
  \centering\raisebox{\dimexpr \topskip-\height}{%
  \includegraphics[width=\textwidth]{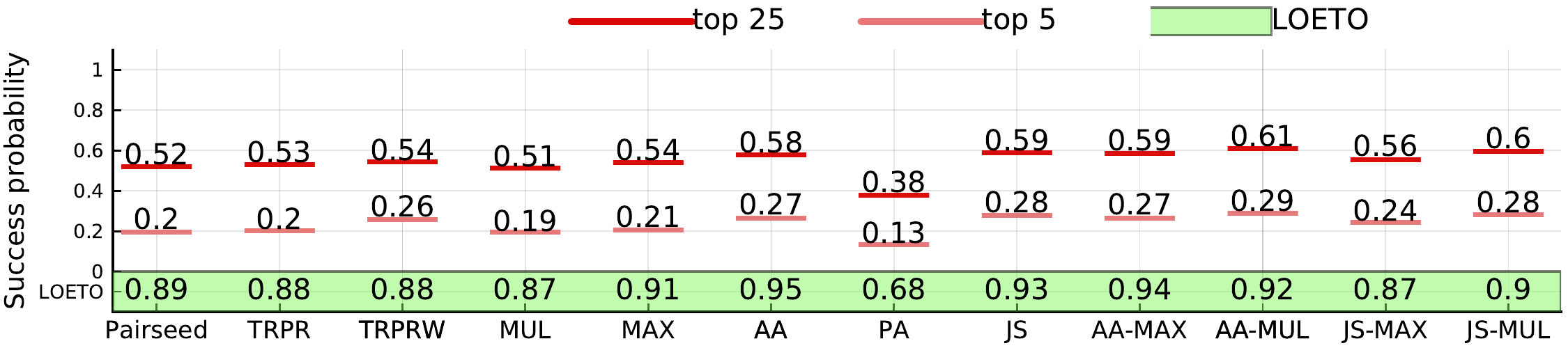}}
\end{minipage}
\begin{minipage}[t]{\dimexpr(0.9\marginparwidth)}
Email-EU (temporal)\\
nodes = 1005\\
edges = 32128\\
triangles = 105461\\
median(nodes to predict) = 1\\
\end{minipage}
\end{fullwidth}
\caption{Binary Mean Value results for the two temporal networks, CollegeMsg and Email-EU. The results on temporal networks are generally worse than the results on static networks, and this can be an indicator that our methods are stronger in predicting missing links rather than future links.}
\label{fig:results_temporal}
\end{figure}

\xhdr{Results} We show the results of all methods in
Figures~\ref{fig:results_bio},~\ref{fig:results_fb},~\ref{fig:results_pa},
and~\ref{fig:results_temporal}. Overall, we notice that the diffusion methods
have more consistency in performance compared to local measures. For instance,
AA-MUL --- which is one of the best performers on some datasets (P-P-Pathways in top
25 and top 5 metrics) --- drops to be one of the worst performers in the top 25
metric on the email dataset. PA is the best performer on the GPA model, but
drops to be the worst performer on all other graphs. In cotrast, TRPRW performs
best on the Ch-Ch-Miner and email datasets but never drops to be one of the worst
methods on any of the datasets. Temporal graphs (CollegeMsg and Email-EU) both
suffered from lower top $k$ scores as compared to static graphs, which suggests
that our methods are possibly stronger in detecting missing links rather than
future links. Upon further investigation on the temporal graphs, we found that
most of the top $k$ predictions were at least two hops away from the seed edges. In the
temporal data, these wedges (length-2 paths) did not close to form triangles
and thus the prediction was incorrect according to the timestamped data. TRPRW
seemed to improve the performance of TRPR in general but did not contribute an
improvement on the email and GPA networks. Upon looking closely at these two
networks, we found that the number of triangles is very small and thus using the
unweighted TRPR version which is close in performance to Pairseed, is more ideal
on datasets that do not contain many triangles.



%% file: 006-linkpred.tex
\section{Back to Standard Link Prediction}
\label{sec:linkpred}

In this section we bring our attention back to the standard link prediction
problem and show how the methods we presented in this paper can also be used to
further enhance standard link prediction. We split our data in the same way to
the previous experiments except that here we use an 80-20 split (often, keeping a higher percentage of the data in the training set produces higher quality results, but in the previous section, we needed to generate more data in the validation set so that we have higher chances of finding paths of length 2 to predict, and thus we increased the size of the testing data by 10 percent). Then, for the top 100 nodes with the largest degree in
the training data, we perform different types of seeded PageRank diffusion for
link prediction on these nodes. This choice of nodes serves the purpose of
identifying nodes that have a higher chance of making connections in the test
data. We measure performance in terms of Area Under the ROC curve (henceforth, AUC score).
Our baseline is single-seeded PageRank.

Our results on pairwise link prediction suggest that multiple seeds with
PageRank-like methods are effective for prediction. Here, we consider four
different multiple-seeding strategies and compare them to single-seeded PageRank
for the classical link prediction problem. We summarize the four new methods in Table~\ref{tab:linkpred}. The
methods \textit{sum}, \textit{max}, and \textit{star-seed} are motivated by the double seeding idea used in the previous sections.

\begin{table}[t]
\caption{description of methods inspired by pairwise link to perform the standard link prediction task}\vspace{-0.5\baselineskip}
\label{tab:linkpred}
\begin{tabularx}{\linewidth}{lX}
  \toprule

  sum\textcolor{plots1}{$\blacktriangle$} & For a certain node $i$, aggregate
  the pair-seeded PageRank results from all edges adjacent to $i$. This is
  equivalent to performing PageRank with a normalized initial vector valued $1$
  at the indices of all the neighbors of $i$, and $\text{degree}(i)$ at index
  $i$. \\ [0.15cm]

  max\textcolor{plots2}{$\bullet$} & This is similar to the previous approach,
  but here, we instead take the element-wise maximum value of the pair-seeded
  PageRank vectors.\\[0.15cm]
  
  star-seed\textcolor{plots3}{\arc} & This is similar to pair-seeded PageRank,
  except that we start PageRank with a normalized initial vector valued $1$ at
  the index of the seed node and all its neighbors.\\ [0.15cm]
  
  TRPR\textcolor{plots4}{$\blacklozenge$} & This uses the same starting vector
  used by star-seed, but instead, applies the TRPR algorithm on
  it.\\
  \bottomrule
\end{tabularx}
\end{table}

We use real-world networks from Section~\ref{sec:results}, and present our
results in Figure~\ref{fig:linkpred}. The scatter plots compare the AUC
score of the neighborhood-based seeding methods to the AUC scores from
single-seeded PageRank. These results suggest that neighborhood-based seeding
is superior to single-seeded PageRank as a link prediction method.

\begin{figure}[t]
\centering
  \begin{minipage}{.75\textwidth}
    \centering
    \includegraphics[width=0.49\linewidth]{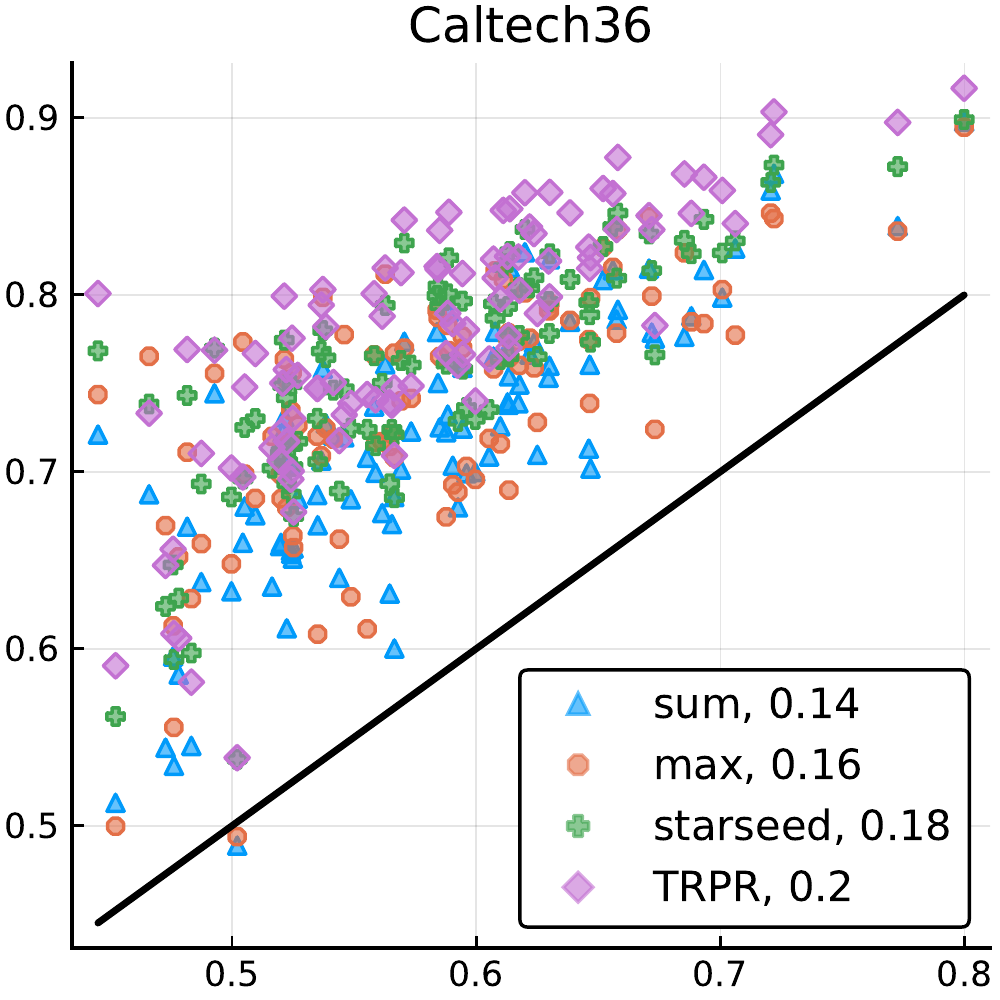}
    \includegraphics[width=0.49\linewidth]{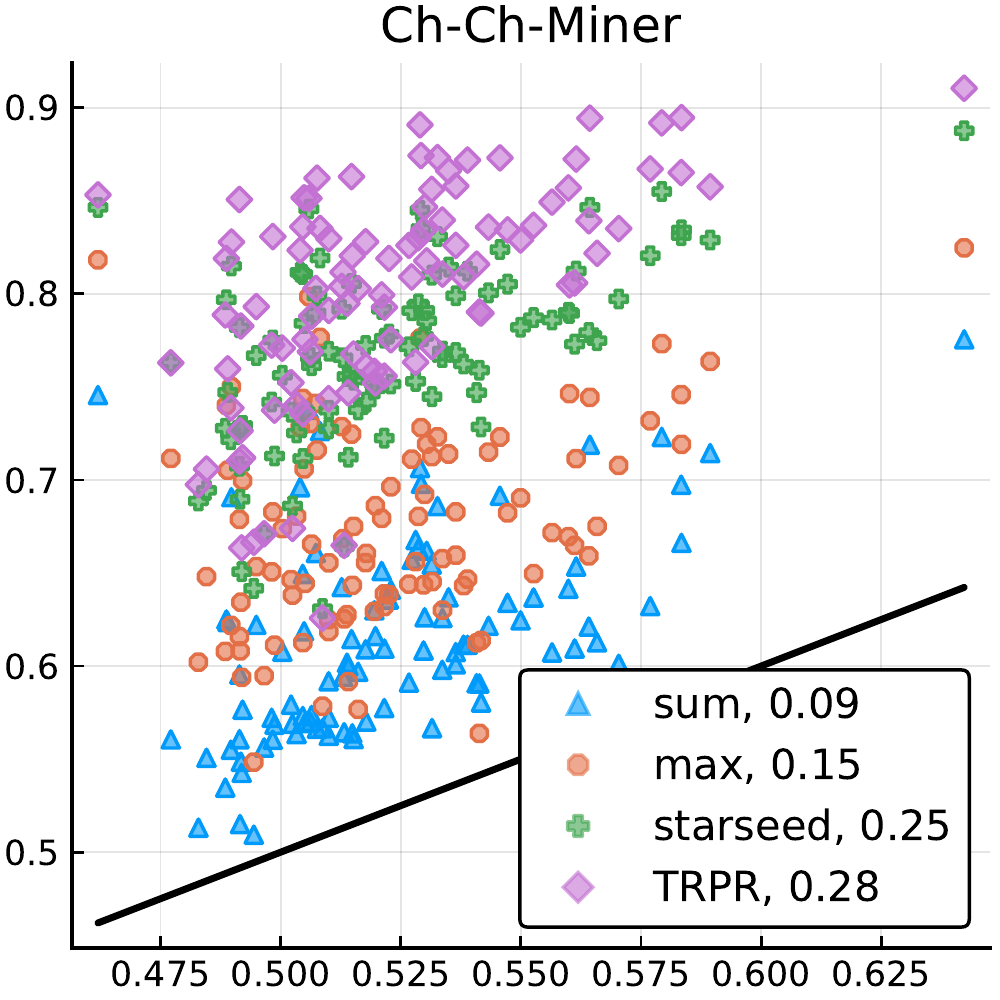}
  \end{minipage}\\%
  \begin{minipage}{.75\textwidth}
    \centering
    \includegraphics[width=0.49\linewidth]{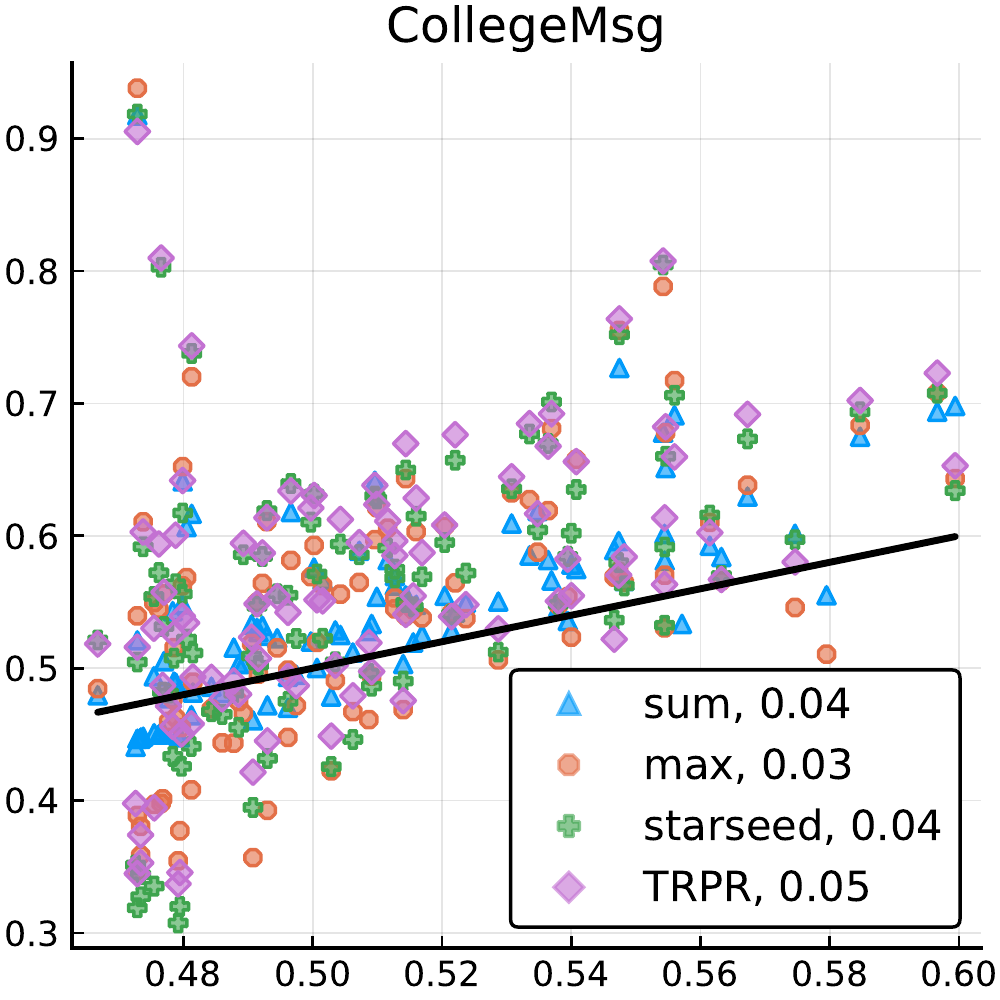}
    \includegraphics[width=0.49\linewidth]{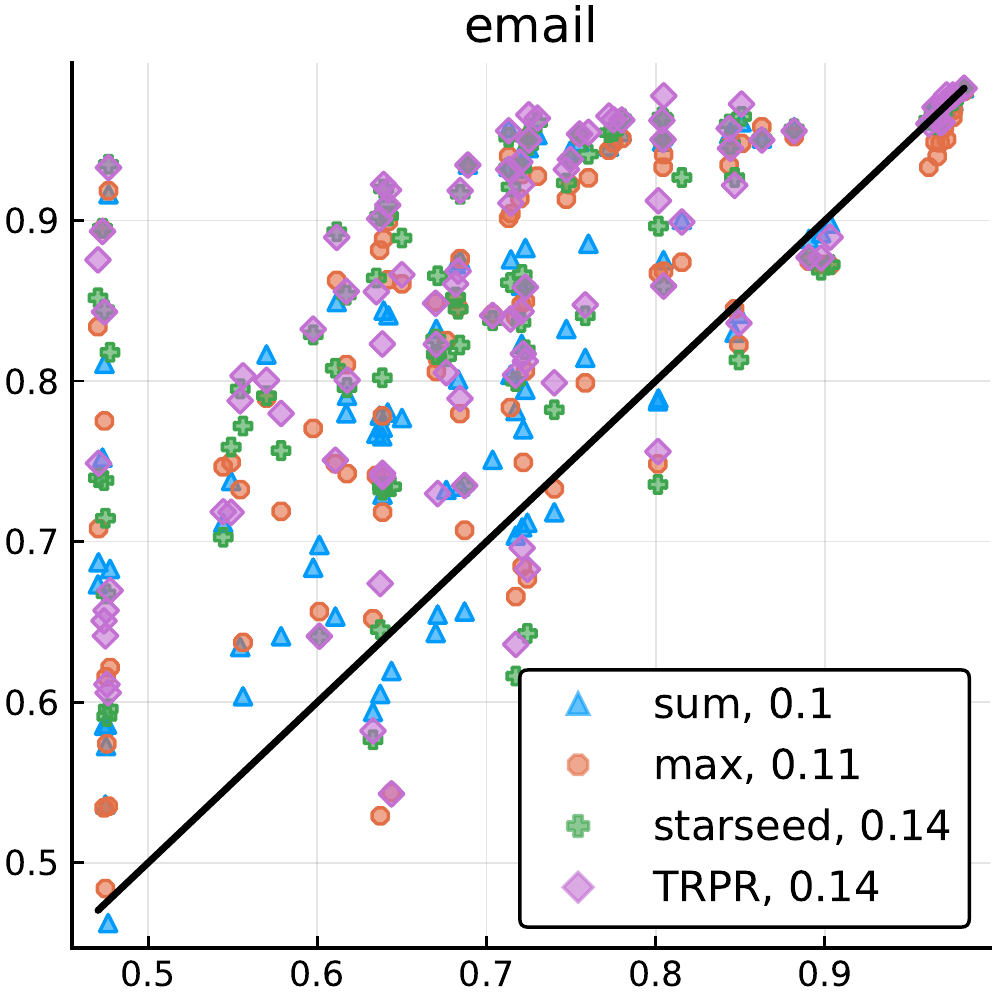}
  \end{minipage}
  \caption{Results of standard link prediction experiment on four
    real-world networks. Each scatter plot shows the link prediction AUC
    results of $100$ experiments of methods inspired by our pairwise link
    prediction proposal with respect to the AUC scores of single-seeded
    PageRank. The solid black line is the plot of $f(x) = x$. Points above
    the line are cases where our proposed methods have superior
    performance to standard single-seeded PageRank.  We see that in most
    cases the four methods outperform the classical seeded PageRank
    method. This study suggests that it is useful to consider a node's
    neighborhood for the purposes of seeding for link prediction with
    PageRank. The values in the legend serve as a summary performance
    measure, which is the average distance to the $f(x) = x$ line. }
  \label{fig:linkpred}
\end{figure}

\hide{
\section{Standard Link Prediction}
\label{sec:linkpred}
In this section we bring our attention back to the standard link prediction problem and we show how the methods we presented in this paper can be used to further enhance link prediction results. We split our data in a similar way to the 80-20 experiments; use $80\%$ of the network for training and $20\%$ for verification. Then, for the nodes corresponding to the top $100$ degree nodes in the training data, we perform the seeded PageRank diffusion for link prediction on these nodes. We evaluate the result with $\mA_{test}$ being our ground truth hidden edges, and treat the resulting AUC scores from seeded PageRank as our baseline in this section. Note that the choice of the top $100$ nodes only serves the purpose of identifying important nodes that have a higher chance of having connections in the testing data. Then, we will use four methods inspired by our pairwise link prediction methods to perform the standard link prediction task. We summarize them below.

\begin{table}
\caption{description of methods inspired by pairwise link to perform the standard link prediction task}\vspace{-0.5\baselineskip}
	\label{tab:linkpred}
	\begin{tabularx}{\linewidth}{lX}
	\toprule
	 sum\textcolor{plots1}{$\blacktriangle$} &  For a certain node $i$, aggregate the pair-seeded PageRank results from all edges adjacent to $i$. This is equivalent to performing PageRank with a normalized initial vector valued $1$ at the indices of all the neighbors of $i$, and $degree(i)$ at index $i$. \\ [0.2cm]
	 max\textcolor{plots2}{$\bullet$} & This is similar to the previous approach, but here, we instead take the element-wise maximum value of the pair-seeded PageRank vectors.\\[0.2cm]
	 
	 star-seed\textcolor{plots3}{\arc} & This is similar to pair-seeded PageRank, except that we start PageRank with a normalized initial vector valued $1$ at the index of the seed node and all its neighbors.\\ [0.2cm]
	 
	 TRPR\textcolor{plots4}{$\blacklozenge$} & This uses the same starting vector used by star-seed, but instead, applies the TRPR algorithm on it.\\

	\bottomrule
\end{tabularx}
\end{table}

%
%
%
%

\xhdr{Neigborhood link prediction} All of these methods fall under the category of using a node's neighborhood for the purpose of predicting its links. We use real world networks used in Section~\ref{sec:real-world}, and present our results in Figure~\ref{fig:linkpred}. The y-axis corresponds to the AUC scores of the neighborhood methods and the x-axis corresponds to the AUC scores from seeded PageRank. These results suggest that neighborhood link prediction methods are superior to seeded PageRank as a link prediction method.

\begin{figure}
	\begin{minipage}{.49\textwidth}
	\centering
		\includegraphics[width=0.49\linewidth]{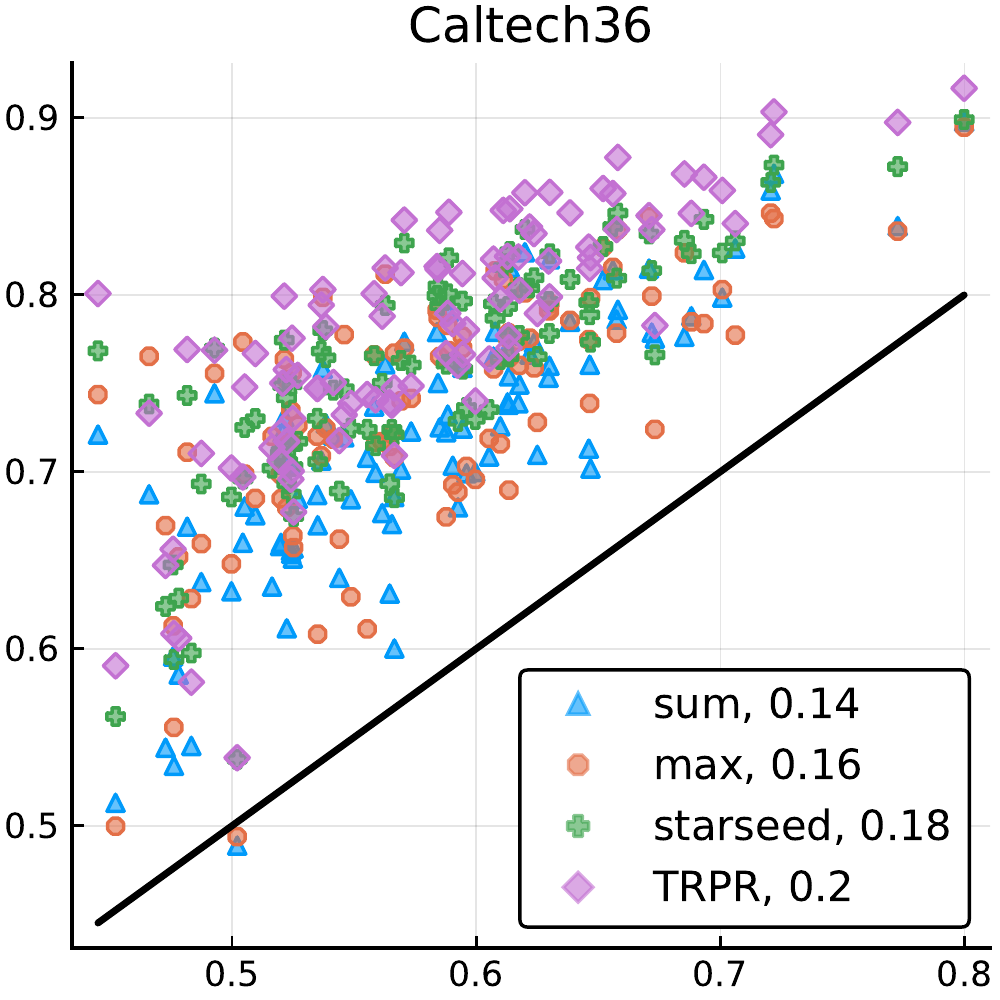}
		\includegraphics[width=0.49\linewidth]{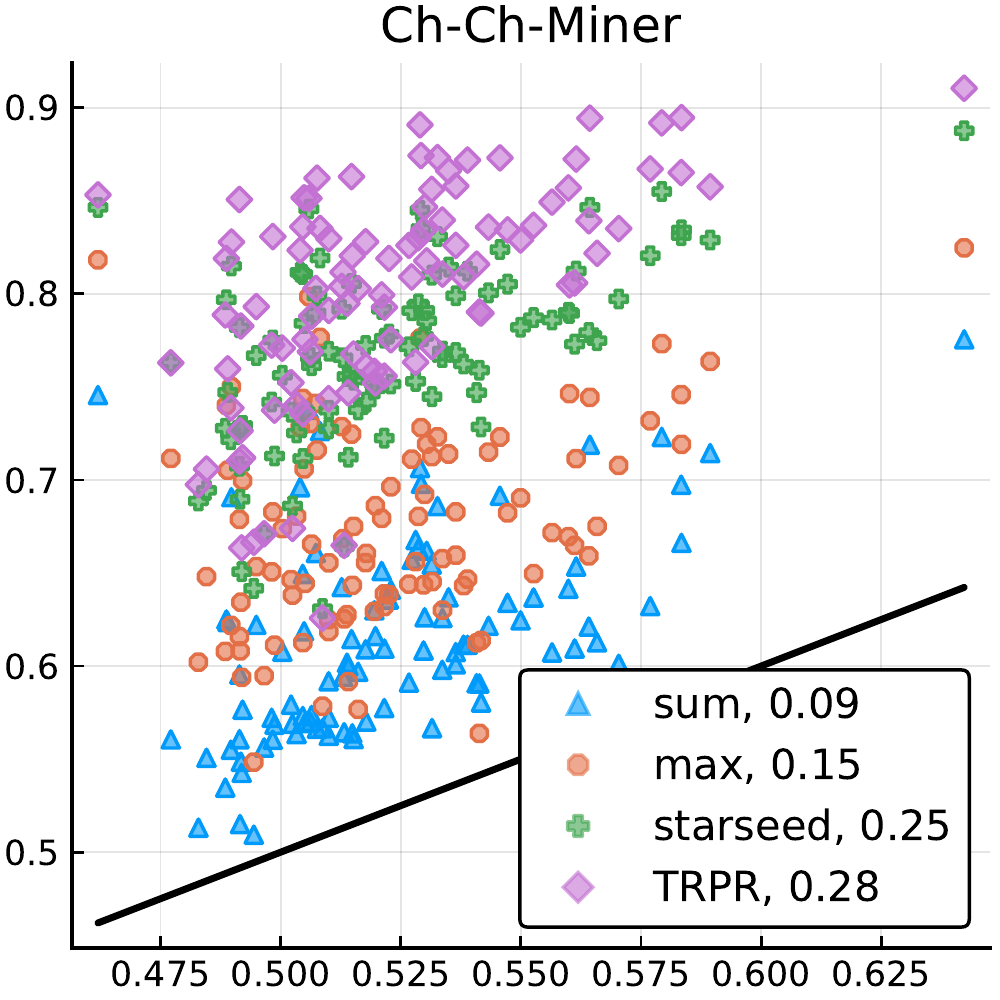}
	\end{minipage}\\%
	\begin{minipage}{.49\textwidth}
	\centering
		\includegraphics[width=0.49\linewidth]{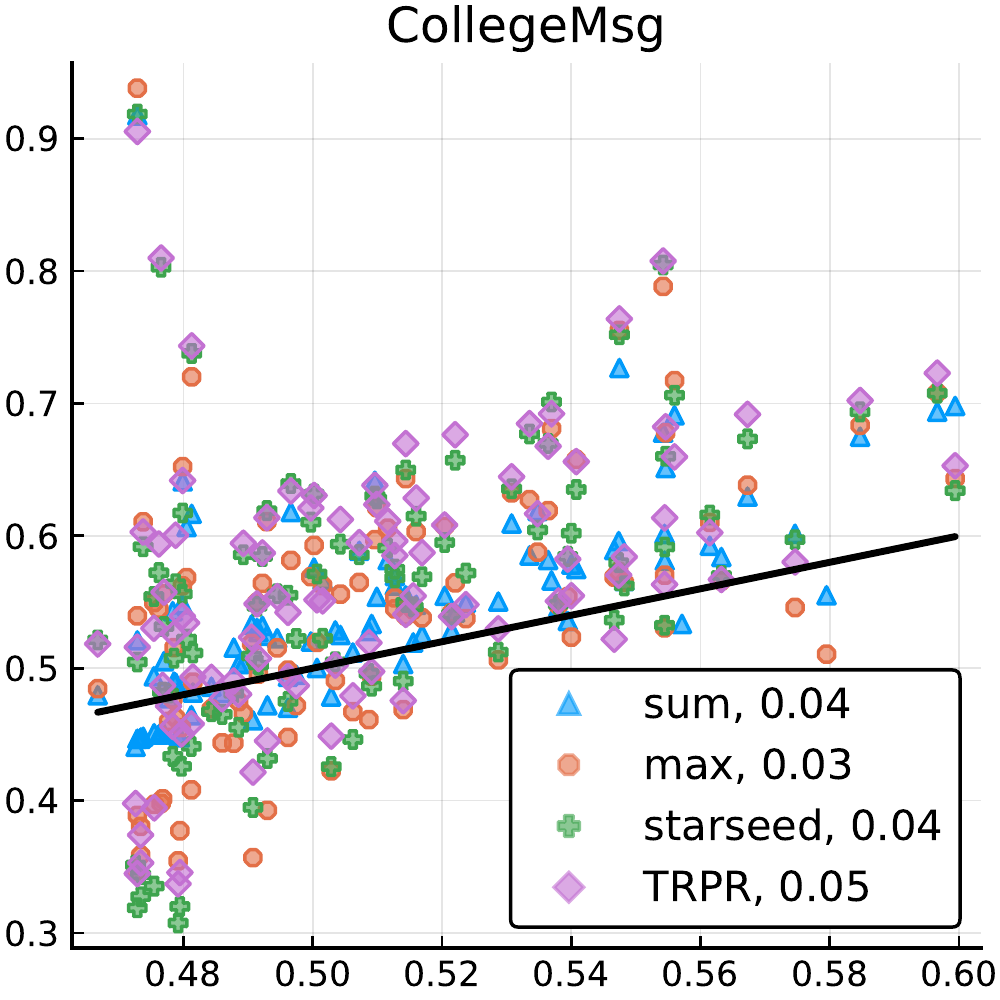}
		\includegraphics[width=0.49\linewidth]{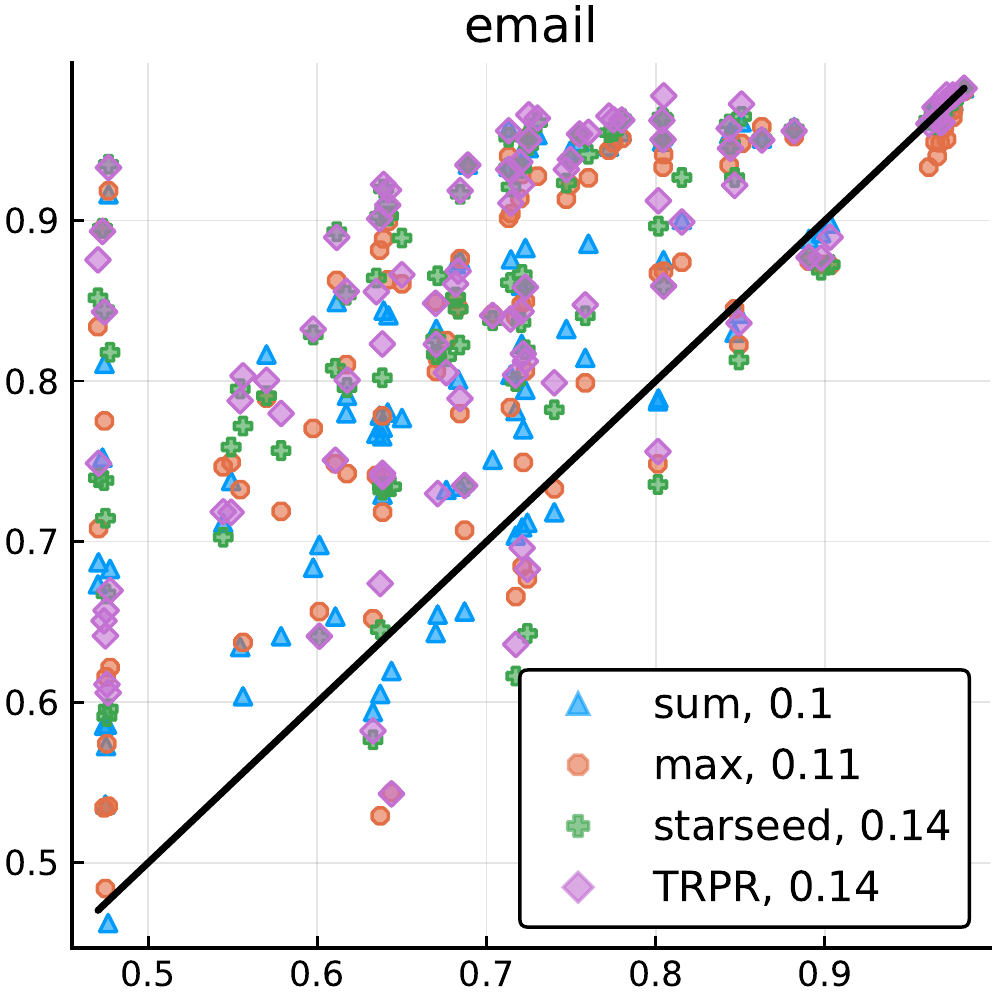}
	\end{minipage}
	\caption{A standard link prediction experiment over 4 real-world graphs. Each scatter plot shows the link prediction AUC results of $100$ experiments of methods inspired by our pairwise link prediction proposal with respect to the AUC scores of seeded PageRank. The solid black line is the plot of $f(x) = x$. We see that in most cases the four methods outperform the classical seeded PageRank method. This study suggests that it is likely useful to consider a node's neighborhood for the purposes of link prediction with PageRank.}
	\label{fig:linkpred}
\end{figure}
}  

%% file: 007-conclusion.tex
\section{Discussion and Future work}
Having a reliable link prediction algorithm is a well-studied research topic due to
its utility in many disciplines. Traditional link prediction methods aim to find
pairs of nodes that are likely to form a link. Here, we have studied a
higher-order version of the problem called pairwise link prediction 
where we predict nodes that are likely to form
a triangle with an edge. We generalized local link-prediction methods 
and we developed two PageRank-based methods for this problem. 
These PageRank-based methods generally remained consistent in behavior on a variety of datasets. Using these results as inspiration,
we then developed multiple-seeding strategies for PageRank in classical link prediction,
which outperform their standard single-seeded counterparts.

TRPR (Triangle Reinforced PageRank) is our new principled method for the task of pairwise link prediction. We demontrated that TRPR is computationally efficient, and demonstarted the implementation details of TRPR can improve on the idealized algorithm by taking advantage of a triangle iterator that avoids building a tensor.
We note that highly efficient implementations of our procedures are possible given their close relationships
with traditional PageRank methods. Scaling to billions of nodes and edges is simply not a
problem given current abilities to compute PageRank (e.g.~\cite{Lofgren-bidirectionalPR-2016}), and 
especially that we have an existing routine to iterate through triangles in a graph quickly.

The space of higher-order prediction problems also has limitless sub-structure.
An alternate problem is to predict an edge that is important when given a single
node. In the future, we intend to extend this work to the latter scenario, and
TRPR can be adapted for this purpose.


%% file: pairseed-arxiv.bbl
\begin{thebibliography}{35}
\providecommand{\natexlab}[1]{#1}
\providecommand{\bibextraformatting}{\relax}
\bibextraformatting

\bibitem[\protect\citeauthoryear{Adamic and
  Adar}{2003}]{Adamic-friends-web-2003}
L.~A. \textsc{Adamic} and E.~\textsc{Adar}.
\newblock \emph{Friends and neighbors on the web}.
\newblock Social Networks, 25~(3), pp. 211 -- 230, 2003.

\bibitem[\protect\citeauthoryear{Agrawal et~al.}{2018}]{Agrawal-PPI-2018}
M.~\textsc{Agrawal}, M.~\textsc{Zitnik}, and J.~\textsc{Leskovec}.
\newblock \emph{Large-scale analysis of disease pathways in the human
  interactome}.
\newblock In \emph{Pacific Symposium on Biocomputing}, p. 111. 2018.

\bibitem[\protect\citeauthoryear{Andersen et~al.}{2006}]{Andersen-PPR-2006}
R.~\textsc{Andersen}, F.~\textsc{Chung}, and K.~\textsc{Lang}.
\newblock \emph{Local graph partitioning using {PageRank} vectors}.
\newblock In \emph{2006 47th Annual {IEEE} Symposium on Foundations of Computer
  Science}. 2006.

\bibitem[\protect\citeauthoryear{Avin et~al.}{2015}]{Avin-GPA-2017}
C.~\textsc{Avin} \textsc{et~al.}
\newblock \emph{Core size and densification in preferential attachment
  networks}.
\newblock In \emph{International Colloquium on Automata, Languages, and
  Programming}, pp. 492--503. 2015.

\bibitem[\protect\citeauthoryear{Backstrom and
  Leskovec}{2011}]{Backstorm-linkprediction-2010}
L.~\textsc{Backstrom} and J.~\textsc{Leskovec}.
\newblock \emph{Supervised random walks: Predicting and recommending links in
  social networks}.
\newblock pp. 635--644. 2011.

\bibitem[\protect\citeauthoryear{Barab{\'a}si and
  Albert}{1999}]{Barbasi-preferential-attachment-1999}
A.-L. \textsc{Barab{\'a}si} and R.~\textsc{Albert}.
\newblock
  \href{https://science.sciencemag.org/content/286/5439/509.full.pdf}{\emph{Emergence
  of scaling in random networks}}.
\newblock Science, 286~(5439), pp. 509--512, 1999.

\bibitem[\protect\citeauthoryear{Benson et~al.}{2017}]{Benson-Spacey-2016}
A.~\textsc{Benson}, D.~F. \textsc{Gleich}, and L.-H. \textsc{Lim}.
\newblock \emph{The spacey random walk: a stochastic process for higher-order
  data}.
\newblock SIAM Review, 59~(2), pp. 321--345, 2017.

\bibitem[\protect\citeauthoryear{Benson et~al.}{2018}]{Benson-simplical-2018}
A.~R. \textsc{Benson}, R.~\textsc{Abebe}, M.~T. \textsc{Schaub},
  A.~\textsc{Jadbabaie}, and J.~\textsc{Kleinberg}.
\newblock \emph{Simplicial closure and higher-order link prediction}.
\newblock Proceedings of the National Academy of Sciences, 2018.

\bibitem[\protect\citeauthoryear{Benson
  et~al.}{2016}]{Benson-organization-2016}
A.~R. \textsc{Benson}, D.~F. \textsc{Gleich}, and J.~\textsc{Leskovec}.
\newblock \emph{Higher-order organization of complex networks}.
\newblock Science, 353~(6295), pp. 163--166, 2016.

\bibitem[\protect\citeauthoryear{Clauset
  et~al.}{2008}]{Clauset-hierarchial-2008}
A.~\textsc{Clauset}, C.~\textsc{Moore}, and M.~E.~J. \textsc{Newman}.
\newblock \href{https://doi.org/10.1038/nature06830}{\emph{Hierarchical
  structure and the prediction of missing links in networks}}.
\newblock Nature, 453, pp. 98 EP --, 2008.

\bibitem[\protect\citeauthoryear{Dave and Hasan}{2019}]{Dave-TCTP-2019}
V.~\textsc{Dave} and M.~\textsc{Hasan}.
\newblock \emph{Triangle completion time prediction using time-conserving
  embedding}.
\newblock ECMLPKDD, 2019.

\bibitem[\protect\citeauthoryear{Easley et~al.}{2010}]{Easley-book-2010}
D.~\textsc{Easley}, J.~\textsc{Kleinberg}, \textsc{et~al.}
\newblock \emph{Networks, crowds, and markets}, Cambridge university press
  Cambridge, 2010.

\bibitem[\protect\citeauthoryear{Eikmeier
  et~al.}{2018}]{Eikmeier-hyperkron-2018}
N.~\textsc{Eikmeier}, A.~S. \textsc{Ramani}, and D.~F. \textsc{Gleich}.
\newblock \emph{The hyperkron graph model for higher-order features}.
\newblock In \emph{{IEEE} International Conference on Data Mining, {ICDM} 2018,
  Singapore, November 17-20, 2018}. 2018.

\bibitem[\protect\citeauthoryear{Gleich}{2015}]{Gleich-PageRank-2015}
D.~F. \textsc{Gleich}.
\newblock \emph{{PageRank} beyond the web}.
\newblock {SIAM} Review, 57~(3), pp. 321--363, 2015.

\bibitem[\protect\citeauthoryear{Gomez-Uribe and
  Hunt}{2015}]{Gomez-Uribe-netflix-2016}
C.~A. \textsc{Gomez-Uribe} and N.~\textsc{Hunt}.
\newblock \emph{The netflix recommender system: Algorithms, business value, and
  innovation}.
\newblock ACM Trans. Manage. Inf. Syst., 6~(4), pp. 13:1--13:19, 2015.

\bibitem[\protect\citeauthoryear{Granovetter}{1977}]{Granovetter-ties-1977}
M.~S. \textsc{Granovetter}.
\newblock \emph{The strength of weak ties}.
\newblock In \emph{Social Networks}, pp. 347--367. Elsevier, 1977.

\bibitem[\protect\citeauthoryear{Guimer\`a
  et~al.}{2003}]{Guimer-selfsimilar-2003}
R.~\textsc{Guimer\`a}, L.~\textsc{Danon}, A.~\textsc{Díaz-Guilera},
  F.~\textsc{Giralt}, and A.~\textsc{Arenas}.
\newblock \emph{Self-similar community structure in a network of human
  interactions}.
\newblock Phys. Rev. E, 68, p. 065103, 2003.

\bibitem[\protect\citeauthoryear{Holland and
  Leinhardt}{1977}]{Holland-structure-1977}
P.~W. \textsc{Holland} and S.~\textsc{Leinhardt}.
\newblock \emph{A method for detecting structure in sociometric data}.
\newblock In \emph{Social Networks}, pp. 411--432. Elsevier, 1977.

\bibitem[\protect\citeauthoryear{Katz}{1953}]{Katz-1953}
L.~\textsc{Katz}.
\newblock \emph{A new status index derived from sociometric analysis}.
\newblock Psychometrika, 18~(1), pp. 39--43, 1953.

\bibitem[\protect\citeauthoryear{Lambiotte
  et~al.}{2019}]{Lambiotte-models-2019}
R.~\textsc{Lambiotte}, M.~\textsc{Rosvall}, and I.~\textsc{Scholtes}.
\newblock \emph{From networks to optimal higher-order models of complex
  systems}.
\newblock Nature Physics, 15~(4), pp. 313--320, 2019.

\bibitem[\protect\citeauthoryear{Liben-Nowell and
  Kleinberg}{2007}]{LibenNowell-link-prediction-2007}
D.~\textsc{Liben-Nowell} and J.~\textsc{Kleinberg}.
\newblock \emph{The link-prediction problem for social networks}.
\newblock Journal of the American Society for Information Science and
  Technology, 58~(7), 2007.

\bibitem[\protect\citeauthoryear{Lin et~al.}{2018}]{Hsu-multimodal-2018}
C.-H. \textsc{Lin}, D.~M. \textsc{Konecki}, M.~\textsc{Liu}, S.~J.
  \textsc{Wilson}, H.~\textsc{Nassar}, A.~D. \textsc{Wilkins}, D.~F.
  \textsc{Gleich}, and O.~\textsc{Lichtarge}.
\newblock \emph{{Multimodal network diffusion predicts future
  disease–gene–chemical associations}}.
\newblock 2018.

\bibitem[\protect\citeauthoryear{Lofgren
  et~al.}{2016}]{Lofgren-bidirectionalPR-2016}
P.~\textsc{Lofgren}, S.~\textsc{Banerjee}, and A.~\textsc{Goel}.
\newblock \emph{Personalized pagerank estimation and search: A bidirectional
  approach}.
\newblock In \emph{Proceedings of the Ninth ACM International Conference on Web
  Search and Data Mining}, pp. 163--172. 2016.

\bibitem[\protect\citeauthoryear{L\"{u} and
  Zhou}{2011}]{Lu-linkpred-survey-2011}
L.~\textsc{L\"{u}} and T.~\textsc{Zhou}.
\newblock \emph{Link prediction in complex networks: A survey}.
\newblock Physica A: Statistical Mechanics and its Applications, 390~(6), 2011.

\bibitem[\protect\citeauthoryear{Milo}{2004}]{Milo-superfamilies-2004}
R.~\textsc{Milo}.
\newblock \emph{Superfamilies of evolved and designed networks}.
\newblock Science, 303~(5663), pp. 1538--1542, 2004.

\bibitem[\protect\citeauthoryear{Milo et~al.}{2002}]{Milo-motif-2002}
R.~\textsc{Milo}, S.~\textsc{Shen-Orr}, S.~\textsc{Itzkovitz},
  N.~\textsc{Kashtan}, D.~\textsc{Chklovskii}, and U.~\textsc{Alon}.
\newblock
  \href{https://science.sciencemag.org/content/298/5594/824.full.pdf}{\emph{Network
  motifs: Simple building blocks of complex networks}}.
\newblock Science, 298~(5594), pp. 824--827, 2002.

\bibitem[\protect\citeauthoryear{Nassar et~al.}{2019}]{Nassar-pairwise-2019}
H.~\textsc{Nassar}, A.~\textsc{Benson}, and D.~F. \textsc{Gleich}.
\newblock \emph{Pairwise link prediction}.
\newblock ASONAM, 2019.

\bibitem[\protect\citeauthoryear{Nassar and Gleich}{2018}]{MatrixNetworks}
H.~\textsc{Nassar} and D.~\textsc{Gleich}.
\newblock \emph{Matrixnetworks.jl}.
\newblock \url{https://github.com/nassarhuda/MatrixNetworks.jl}, 2018.

\bibitem[\protect\citeauthoryear{Newman}{2001}]{Newman-growing-networks-2001}
M.~E.~J. \textsc{Newman}.
\newblock \emph{Clustering and preferential attachment in growing networks}.
\newblock Phys. Rev. E, 64, p. 025102, 2001.

\bibitem[\protect\citeauthoryear{Page et~al.}{1999}]{Page-PageRank-1999}
L.~\textsc{Page}, S.~\textsc{Brin}, R.~\textsc{Motwani}, and
  T.~\textsc{Winograd}.
\newblock \emph{The pagerank citation ranking: Bringing order to the web.}
\newblock Technical Report 1999-66, Stanford InfoLab, 1999.
\newblock Previous number = SIDL-WP-1999-0120.

\bibitem[\protect\citeauthoryear{Panzarasa
  et~al.}{2009}]{Panzarasa-college-2009}
P.~\textsc{Panzarasa}, T.~\textsc{Opsahl}, and K.~M. \textsc{Carley}.
\newblock \emph{Patterns and dynamics of users{\textquotesingle} behavior and
  interaction: Network analysis of an online community}.
\newblock Journal of the American Society for Information Science and
  Technology, 60~(5), pp. 911--932, 2009.

\bibitem[\protect\citeauthoryear{Rapoport}{1953}]{Rapoport-transitivity-1953}
A.~\textsc{Rapoport}.
\newblock \emph{Spread of information through a population with
  socio-structural bias: I. assumption of transitivity}.
\newblock The Bulletin of Mathematical Biophysics, 15~(4), pp. 523--533, 1953.

\bibitem[\protect\citeauthoryear{{Stanford SNAP Group}}{2017}]{biosnapnets}
\textsc{{Stanford SNAP Group}}.
\newblock \emph{Miner: Gigascale multimodal biological network}.
\newblock \url{https://github.com/snap-stanford/miner-data}, 2017.

\bibitem[\protect\citeauthoryear{Traud et~al.}{2011}]{Traud-facebook100-2011}
A.~L. \textsc{Traud}, P.~J. \textsc{Mucha}, and M.~A. \textsc{Porter}.
\newblock \emph{Social structure of facebook networks}.
\newblock CoRR, abs/1102.2166, 2011.
\newblock \href {http://arxiv.org/abs/1102.2166}
  {\normalcolor\path{arXiv:1102.2166}}.

\bibitem[\protect\citeauthoryear{Wishart et~al.}{2017}]{Wishart-drugbank-2017}
D.~S. \textsc{Wishart} \textsc{et~al.}
\newblock \emph{{DrugBank} 5.0: a major update to the {DrugBank} database for
  2018}.
\newblock Nucleic Acids Research, 2017.

\end{thebibliography}
